\documentclass[12pt]{iopart}

\usepackage{iopams}
\usepackage{graphicx}
\usepackage{mathrsfs}
\usepackage{color}
\usepackage[table]{xcolor}
\usepackage{hyperref}
\hypersetup{colorlinks, citecolor=black, filecolor=black, linkcolor=black, urlcolor=black}

\expandafter\let\csname equation*\endcsname\relax
\expandafter\let\csname endequation*\endcsname\relax
\usepackage{amsmath}

\usepackage{dcolumn}% Align table columns on decimal point
\usepackage{bm}% bold math
\usepackage{setspace}% shrink spacing between lines

\usepackage[export]{adjustbox}

\usepackage{empheq}
\usepackage[most]{tcolorbox}

\definecolor{grayish}{RGB}{230,230,230}

\newcommand{\refEq}[1] {(\ref{#1})}

\newcommand{\Sin}[1]{\ensuremath{\sin \left( #1 \right)}}
\newcommand{\Cos}[1]{\ensuremath{\cos \left( #1 \right)}}

\newcommand{\Nabla}{\ensuremath{\vec{\nabla}}}
\newcommand{\romanNum}[1]{\uppercase\expandafter{\romannumeral#1}}

\newcommand{\tensor}[1]{\overset{\text{\scriptsize$\leftrightarrow$}}{#1}}

\begin{document}

\title{Local gyrokinetic simulations of tokamaks with non-uniform magnetic shear}

\author{Justin Ball and Stephan Brunner}

\address{Ecole Polytechnique F\'{e}d\'{e}rale de Lausanne (EPFL), Swiss Plasma Center (SPC), CH-1015 Lausanne, Switzerland}

\ead{Justin.Ball@epfl.ch}

\begin{abstract}

In this work, we modify the standard flux tube simulation domain to include arbitrary ion gyroradius-scale variation in the radial profile of the safety factor. To determine how to appropriately include such a modification, we add a strong ion gyroradius-scale source (inspired by electron cyclotron current drive) to the Fokker-Planck equation, then perform a multi-scale analysis that distinguishes the fast electrons driven by the source from the slow bulk thermal electrons. This allows us to systematically derive the needed changes to the gyrokinetic model. We find new terms that adjust the ion and electron parallel streaming to be along the modified field lines. These terms have been successfully implemented in a gyrokinetic code (while retaining the typical Fourier representation), which enables flux tube studies of non-monotonic safety factor profiles and the associated profile shearing. As an illustrative example, we investigate tokamaks with positive versus negative triangularity plasma shaping and find that the importance of profile shearing is not significantly affected by the change in shape.

\end{abstract}

%%===================================================%
%%===================================================%
\section{Introduction}
\label{sec:intro}
%%===================================================%
%%===================================================%

Gyrokinetic simulations are the highest-fidelity tool used to study turbulence in magnetic fusion devices. While they typically require a supercomputer, nonlinear gyrokinetic simulations can give quantitatively accurate predictions of energy transport and many other statistical properties of turbulence \cite {JenkoGENEvalidation2013, GoerlerGENEvalidation2014}. This capability is invaluable for evaluating the performance of proposed designs, interpreting measurements on existing experiments, and improving our understanding of the physics of plasmas.

The gyrokinetic model is a result of a formal asymptotic expansion of the Fokker-Planck kinetic equation in $\rho_{\ast} \equiv \rho_{i} / a$, the ratio of the ion gyroradius $\rho_{i}$ to the tokamak minor radius $a$. All quantities are given a size with respect to $\rho_{\ast}$ and, by proceeding order by order, the equations governing turbulence, neoclassical physics, MHD equilibrium, and transport can be systematically derived \cite{AbelGyrokineticsDeriv2012}. This expansion explicitly separates the time and space scales of the turbulence from those of the background plasma equilibrium --- the amplitude of turbulent fluctuations are assumed to be a factor of $\rho_{\ast}$ smaller than the background equilibrium, while the timescale of the turbulence is a factor of $\rho_{\ast}^{2}$ faster than the equilibrium. Similarly, in the directions perpendicular to the magnetic field the turbulent eddies have a spatial size comparable to $\rho_{i}$, a factor of $\rho_{\ast}$ smaller than the background quantities, which vary over the scale of $a$. On the other hand, parallel to the magnetic field, the eddies are very extended and have a length comparable to $a$.

Because of the separation of scales at the heart of gyrokinetics and the anisotropy of turbulence, it is natural to use a ``flux tube'' \cite{BeerBallooningCoordinates1995} --- a simulation domain that is very elongated along the magnetic field line and narrow in the perpendicular directions. The perpendicular box widths are measured in $\rho_{i}$, while the parallel length is measured in number of poloidal turns around the tokamak cross-section. Thus, all the radial profiles of equilibrium parameters (i.e. the density, flow, temperature, safety factor, and flux surface shape) can be Taylor expanded to first order around the flux surface at the center of the domain to reduce their variation to a simple linear dependence. For this reason, simulations employing a flux tube are called ``local.'' Additional terms in the Taylor expansion (e.g. the curvature of the profiles) are referred to as ``profile shearing'' and are generally neglected in local gyrokinetics because they are small in $\rho_{\ast} \ll 1$. This means the safety factor (i.e. the number of {\it toroidal} turns a magnetic field line makes per {\it poloidal} turn of the tokamak) profile is parameterized in local simulations by just two scalar quantities: the value of the safety factor at the center of the domain $q_{0}$ and the radial derivative of the safety factor $\hat{s} \equiv (r_{0}/q_{0}) dq / dr$. Specifically, it becomes
\begin{align}
  q(r) = q_{0} \left( 1 + \hat{s} \frac{r - r_{0}}{r_{0}} \right) , \label{eq:qForm}
\end{align}
where $r$ is a flux surface label and $r_{0}$ is its value at the center of the domain. Recently there have been several efforts to include profile shearing in temperature and density \cite{PueschelProfCorrugations2013, CandySpectralTemp2020, StOngeGlobalStella2022}, flow \cite{PueschelProfCorrugations2013, CandySpectralFlow2018}, flux surface shape \cite{StOngeGlobalStella2022}, and the safety factor \cite{StOngeGlobalStella2022}. Importantly for this work, the approach of \cite{StOngeGlobalStella2022} requires the safety factor profile to be monotonic and assumes it varies over a much longer spatial scale than that of the turbulence. Additionally, because flux tubes model an asymptotically narrow range of flux surfaces, it becomes appropriate to apply periodic boundary conditions in the radial direction, in addition to the binormal and parallel directions \cite{BeerBallooningCoordinates1995}. Thus, a Fourier representation is typically employed in the directions perpendicular to the magnetic field line, which has several advantages (e.g. the 3/2 rule can be employed to prevent aliasing issues \cite{Orszag3/2rule1971,ToldShiftedMetric2010} and the gyroaverage becomes a simple multiplication by a Bessel function \cite{CattoLinearizedGyrokinetics1978}).

It is also common to perform gyrokinetic simulations using a ``global'' domain \cite{CandyGYRO2003, IdomuraGT5D2008, GoerlerGlobalGENE2011, GrandgirardGYSELA2016, LantiORB52020}, which spans a large fraction of the tokamak cross-section. While global simulations generally still solve the same gyrokinetic model that results from the asymptotic expansion in $\rho_{\ast} \ll 1$, they retain the full radial profiles of the background quantities. Thus, such simulations include profile shearing as well as other effects that are formally small in $\rho_{\ast}$. While global simulations cannot be claimed to be more accurate as they don't contain {\it all} terms that appear to next order in the $\rho_{\ast}$ expansion \cite{Calvo2ndOrderPartTwo2012}, including {\it some} formally small terms can still be useful. For example, comparing global and local simulations gives information about what numerical value of $\rho_{\ast}$ is needed for the asymptotic expansion to be valid \cite{McMillanSystemSizeScaling2010}.

Unfortunately, global simulations are substantially more computationally expensive and less robust due to their greater complexity. They typically consider significantly larger physical domains, which include a wider range of plasma conditions that must be properly resolved. Additionally, because they include both turbulent and transport timescales, they must include sources and sinks of particles and energy \cite{CandyGYRO2003, GoerlerGlobalGENE2011} to prevent turbulence from slowly flattening the driving gradients. Lastly, due to the complexities of the plasma edge, the appropriate radial boundary conditions are unclear \cite{GeraldiniSheath2018}. Typically, Dirichlet boundary conditions are used together with buffer regions \cite{CandyGYRO2003, GoerlerGlobalGENE2011}, but then the results must be tested to ensure that they are not contingent on such an artificial boundary condition. The boundary conditions, together with the explicit radial dependence of geometric coefficients in the equations, typically prevent global simulations from employing a Fourier representation in the radial direction.

In this work, we will bridge the separation of scales in a novel way --- we will enable the flux tube simulation domain to model ion gyroradius-scale variation in the radial profile of the magnetic shear. Specifically, we aim to model variation with a scale that is tens to hundreds of gyroradii, yet still asymptotically smaller than the tokamak minor radius. Importantly, this can be done in a computationally efficient manner, while retaining the practical advantages of the flux tube. In particular, sources/sinks are not required, periodic radial boundary conditions can be applied, and the perpendicular spatial grid can still be discretized by a Fourier decomposition. This will enable reliable local studies of profile shearing in the safety factor profile. Additionally, the model also includes shearing in the steady-state profiles of temperature, density, and flow because these will invariably adjust during the simulation to be consistent with the externally imposed safety factor profile and the absence of sources/sinks. One interesting potential application is reversed magnetic shear profiles, which have been found to enable internal transport barrier formation \cite{JoffrinITBintegerq2003, WaltzShearLayers2006}. Such profiles include a point with $\hat{s}=0$, which {\it may} make profile curvature particularly important. As discussed in the conclusions, other potential applications include reducing the computation cost of simulating very low but finite values of magnetic shear, investigating turbulent broadening of current drive sources, and performing self-consistent studies of profile shearing in the density, flow, and temperature profiles.

In section \ref{sec:orderings}, we will explain how ion gyroradius-scale radial variation might arise in experimental magnetic shear profiles and explain how it can fit into the standard gyrokinetic asymptotic ordering. Then, in section \ref{sec:deriv} we will derive the equations for a flux tube with non-uniform magnetic shear and implement them in the local gyrokinetic code GENE \cite{JenkoGENE2000}. In section \ref{sec:benchmark}, we will benchmark the code against linear analytic results as well as conventional nonlinear flux tube simulations. Next, in section \ref{sec:example} we will present an example application to illustrate its potential uses: comparing the importance of profile shearing in positive and negative triangularity tokamaks. Finally, in section \ref{sec:conclusions} we will provide some concluding remarks.

%%===================================================%
%%===================================================%
\section{Physical origins and orderings}
\label{sec:orderings}
%%===================================================%
%%===================================================%

There are at least two mechanisms that could plausibly create ion gyroradius-scale variation in the radial profile of the magnetic shear --- Electron Cyclotron Current Drive (ECCD) and the bootstrap current. ECCD is one of the primary steady-state methods used to drive the toroidal plasma current \cite{PraterECCD2004} (which gives rise to the poloidal magnetic field and thereby determines the safety factor and magnetic shear profiles). It is one of only two non-inductive current drive systems planned for the initial operating phase of ITER \cite{PolevoiITERcurrentDrive2020}. Its widespread use is largely due to its ability to provide highly localized current drive, thereby enabling fine control over the safety factor profile \cite{PraterECCD2004}. In fact, the skin depth of radio-frequency waves around the electron cyclotron resonance can be just $\sim 10 \rho_{i}$ \cite{PraterECCD2004, GoodmanTCVeccd2005, MaraschekAUGeccd2005, HarveyITEReccd2001}, similar to the typical size of turbulent eddies. Thus, an ECCD system can be sufficiently localized to create a gyroradius-scale source of toroidal current. Accordingly, there is evidence that ion gyroradius-scale structure can be created experimentally (e.g. figure 15 of \cite{ZuccaQprofs2008}, figure 5.9 of \cite{ArgomedoQprof2014}), although the magnetic shear profile is difficult to measure directly. Alternatively, we note that the bootstrap current arising from transport barriers can also have quite small-scale radial variation (e.g. figure 5 of \cite{GroebnerPedCurrent1998}, figure 3.19 of \cite{GibsonPedCurrent2021}, figure 1 of \cite{LapillonneMagShear2011}). While we are not seeking to rigorously model either of these sources of current drive, we will use the characteristics of ECCD to motivate approximations to arrive at a plausible and internally consistent simplified model that features non-uniform magnetic shear.

To show that a magnetic shear profile that varies across a flux tube and is fixed in time can be made consistent with the gyrokinetic orderings, we will follow the derivation in \cite{AbelGyrokineticsDeriv2012}, but add a new source term $\tilde{S}^{Ip}_{e} (r)$. Prior to the asymptotic expansion, this source term would appear on the right side of the Fokker-Planck kinetic equation (i.e. equation (1) of \cite{AbelGyrokineticsDeriv2012}). We will define this term to be a source of toroidal plasma current $I_{p}$ that is carried by the electrons, varies radially on the ion gyroradius-scale (i.e. $d \tilde{S}^{Ip}_{e} / dr \sim \tilde{S}^{Ip}_{e}/\rho_{i}$), and averages to zero over the turbulent spatial scale in the radial direction (i.e. $\langle \tilde{S}^{Ip}_{e} (r) \rangle_{\text{turb}}=0$). We want to choose the strength of $\tilde{S}^{Ip}_{e} (r)$ such that the modulation it creates in the magnetic shear $\tilde{s} (r)$ competes against the standard magnetic shear $\hat{s}$ of the background linear safety factor profile (i.e. $\tilde{s} (r) \sim \hat{s}$). Regardless, because of its small radial scale, this causes a negligible modification to the safety factor (i.e. $(q_{0}/r_{0}) \int dr \tilde{s}(r) \ll q_{0}$). This is analogous to other background profiles, like temperature or density, for which the turbulent fluctuations are strong enough to locally flatten the background gradients, but not to modify the background value itself. To accomplish this, we will order the strength of the source as $\tilde{S}^{Ip}_{e} \sim \rho_{\ast} \omega F_{s}$, where $F_{s}$ is the background distribution function and $\omega$ is the frequency of the turbulence. 

Consequently, when performing the asymptotic expansion in $\rho_{\ast} \ll 1$, the new current drive source term will first appear at the order of neoclassical theory and gyrokinetics (i.e. at $O(\rho_{\ast} \omega F_{s})$). Since the source averages to zero over the turbulent spatial scale, it does not appear in the neoclassical drift kinetic equation, the Grad-Shafranov equation, nor the evolution equation for the {\it mean} magnetic field. In fact, the only place it appears is on the right side of the electron gyrokinetic equation. To next order (i.e. at $O(\rho_{\ast}^{2} \omega F_{s})$), it appears in the transport equations. There its gyroradius-scale contributions still average to zero, but it should be pointed out that it could have an equilibrium-scale contribution. Specifically if the source depends on the distribution function, an equilibrium-scale contribution could arise from a beating between the spatial variation of the source and the spatial variation of the turbulent portion of the distribution function. This would have to be included in the transport modeling, but it is outside the scope of this work as we are exclusively concerned with the gyrokinetic calculation.

%%===================================================%
%%===================================================%
\section{Analytic gyrokinetic derivation}
\label{sec:deriv}
%%===================================================%
%%===================================================%

The starting point of our derivation is the real-space electromagnetic gyrokinetic model \cite{ParraUpDownSym2011} with the addition of the new source term $\tilde{S}^{Ip}_{e} (r)$. The kinetic equation is given by
\begin{align}
  \frac{\partial h_{s}}{\partial t} &+ v_{||} \hat{b} \cdot \Nabla h_{s} + \left( \vec{c}_{\kappa} v_{||}^{2} + \vec{c}_{\nabla B} \mu \right) \cdot \Nabla h_{s} + c_{a||} \mu \frac{\partial h_{s}}{\partial v_{||}} - \sum_{s'} \left\langle C^{L}_{s s'} \right\rangle_{\varphi} \label{eq:GKeq} \\
  &- \frac{1}{B} \left( \Nabla h_{s} \times \Nabla \langle \chi \rangle_{\varphi} \right) \cdot \hat{b} = \frac{Z_{s} e F_{Ms}}{T_{s}} \frac{\partial \langle \chi \rangle_{\varphi}}{\partial t} + \frac{1}{B} \left( \Nabla F_{Ms} \times \Nabla \langle \chi \rangle_{\varphi} \right) \cdot \hat{b} + \left\langle \tilde{S}^{Ip}_{s} \right\rangle_{\varphi} , \nonumber
\end{align}
where $\tilde{S}^{Ip}_{s}$ is only non-zero for electrons. The unknowns are the non-adiabatic portion $h_{s} = \delta f_{s} + (Z_{s} e \phi/T_{s}) F_{Ms}$ of the perturbed turbulent distribution function $\delta f_{s}$ and the gyroaveraged fluctuating generalized potential $\langle \chi \rangle_{\varphi} \equiv \langle \phi \rangle_{\varphi} - v_{||} \langle A_{||} \rangle_{\varphi} - \langle \vec{v}_{\perp} \cdot \vec{A}_{\perp} \rangle_{\varphi}$, comprising the electrostatic potential $\phi$ and the fluctuating magnetic field $\delta \vec{B} = \Nabla \times \vec{A}$ (which is determined through the magnetic vector potential  $\vec{A}$). The distribution function $h_{s}$ is a function of the guiding center position $\vec{X}$, parallel velocity $v_{||}$, magnetic moment $\mu \equiv v_{\perp}^{2}/(2 B)$, and time $t$. The fields depend on the particle position $\vec{x} = \vec{X} + \vec{\rho}_{s}$, but the gyroaverage $\left\langle \ldots \right\rangle_{\varphi} \equiv (2 \pi)^{-1} \oint_{0}^{2\pi} |_{\vec{X}} d \varphi ( \ldots )$ over the gyroangle $\varphi$ is taken holding the guiding center position constant. The subscripts $||$ and $\perp$ refer to the vector components parallel and perpendicular to the direction of the background magnetic field $\hat{b} \equiv \vec{B}/B$ respectively, where $\vec{B}$ is the background magnetic field and $B$ is its magnitude. The curvature drift $\vec{c}_{\kappa} v_{||}^{2} = (v_{||}^{2}/\Omega_{s}) \hat{b} \times ( \hat{b} \cdot \Nabla \hat{b} )$ and the grad-B drift  $\vec{c}_{\nabla B} \mu = (\mu/\Omega_{s}) \hat{b} \times \Nabla B$ are written in a form to stress their velocity dependences, as is the parallel acceleration $c_{a||} \mu = - \mu \hat{b} \cdot \Nabla B$ (i.e. the mirror term). Collisions are included through the linearized collision operator $C^{L}_{s s'}$ between species $s$ and $s'$. The particle charge number of species $s$ is indicated by $Z_{s}$, $e$ is the elementary charge, $F_{Ms}$ is the unshifted Maxwellian background distribution function with a temperature $T_{s}$ and number density $n_{s}$, $\vec{\rho}_{s} = \hat{b} \times \vec{v}_{\perp} / \Omega_{s}$ is the gyroradius vector, $\vec{v}_{\perp}$ is the perpendicular velocity, $\Omega_{s} = Z_{s} e B/m_{s}$ is the gyrofrequency, and $m_{s}$ is the particle mass. Note that we have ignored plasma rotation for simplicity. The model is closed through the field equations of quasineutrality and the parallel and perpendicular components of Ampere's law, given by
\begin{align}
  \phi &= \left( \sum_{s} \frac{Z_{s}^{2} e^{2} n_{s}}{T_{s}} \right)^{-1} \sum_{s} Z_{s} e B \int_{-\infty}^{\infty} dv_{||} \int_{0}^{\infty} d \mu \left. \oint_{0}^{2 \pi} \right|_{\vec{x}} d \varphi ~ h_{s} \left( \vec{x} - \vec{\rho}_{s} \right) \label{eq:fieldPhi} \\
  - \nabla_{\perp}^{2} A_{||} &= \mu_{0} \sum_{s} Z_{s} e B \int_{-\infty}^{\infty} dv_{||} \int_{0}^{\infty} d \mu \left. \oint_{0}^{2 \pi} \right|_{\vec{x}} d \varphi ~ v_{||} h_{s} \left( \vec{x} - \vec{\rho}_{s} \right) \label{eq:fieldApar} \\
  \Nabla \delta B_{||} \times \hat{b} &= \mu_{0} \sum_{s} Z_{s} e B \int_{-\infty}^{\infty} dv_{||} \int_{0}^{\infty} d \mu \left. \oint_{0}^{2 \pi} \right|_{\vec{x}} d \varphi ~ \vec{v}_{\perp} h_{s} \left( \vec{x} - \vec{\rho}_{s} \right) \label{eq:fieldBpar}
\end{align}
respectively, where $\mu_{0}$ is the permeability of free space. Throughout this paper, for brevity, we will often omit some of the functional dependencies of quantities (e.g. the velocity dependence of the distribution function or the poloidal dependence of the source $\tilde{S}^{Ip}_{e}$) if they are not pertinent.

Rather than implementing a realistic current drive source in the gyrokinetic equation \cite{DonnelECCD2022}, we will carry out a subsidiary asymptotic expansion inspired by properties of ECCD. This will produce a reasonable and internally consistent simplified model that features non-uniform magnetic shear. Specifically, we will perform multi-scale analysis \cite{BenderAdvMathMethods1999} to distinguish the standard electron thermal speed $v_{the}$ from the characteristic speed of the current drive source $v_{f}$, which we will consider to be asymptotically fast. {\it This is the central approximation of the derivation and we believe it to be reasonable, given that ECCD is thought to act on electrons with speeds several times larger than $v_{the}$} \cite{PraterECCD2004}. To execute the multi-scale analysis, we substitute $v_{||} \rightarrow v_{||} + v_{||f}$ and $\mu \rightarrow \mu + \mu_{f}$ for electron dynamics. Here $v_{||f} \equiv \epsilon v_{||}$ and $\mu_{f} \equiv \epsilon \mu$ are the new fast velocity coordinates and $\epsilon \equiv v_{the}/v_{f} \ll 1$ is the small parameter of our subsidiary expansion (but $\epsilon$ is still taken as $O(1)$ in the context of the primary $\rho_{\ast} \ll 1$ expansion). The definitions of $v_{||f}$ and $\mu_{f}$ were chosen to be appropriate for parameterizing fast velocity scales, while $v_{||}$ and $\mu$ parameterize thermal velocity scales. Accordingly, we adopt the orderings $v_{||} \sim v_{the}$, $v_{||f} \sim v_{f}$, $\mu \sim v_{the}^{2}/B$, and $\mu_{f} \sim v_{the} v_{f}/B$. Note, in particular, the ordering of $\mu_{f}$, which is suitable if the source $\tilde{S}^{Ip}_{s}$ injects a similar amount of momentum in the parallel and perpendicular components of the electron velocity. Moreover, we will see that such an ordering for $\mu_{f}$ is needed to allow the fast and thermal contributions to compete in both components of Ampere's law (i.e. equations \refEq{eq:fieldApar} and \refEq{eq:fieldBpar}). While we have taken $v_{the} \ll v_{f}$, we still let $v_{f} \ll (m_{i}/m_{e}) v_{thi}$, so that the gyroradius of the fast electrons remains much smaller than the thermal ion gyroradius. Next, we expand $h_{s} = h_{s0} (v_{||}, \mu, v_{||f}, \mu_{f}) + h_{s1} (v_{||}, \mu, v_{||f}, \mu_{f}) + \ldots$ as well as the fields within $\chi$, where the numerical subscript indicates the quantity's relative size in $\epsilon \ll 1$. We carefully choose our ordering in $\epsilon$ for the current drive source $\tilde{S}^{Ip}_{e} ( v_{||f}, \mu_{f} ) \sim \epsilon h_{e} v_{the}/a$, which will enable it to affect the turbulent dynamics, but not to dominate. Note that the source is assumed to vary only on the fast velocity scale. Lastly, we will assume that $\tilde{S}^{Ip}_{e} ( v_{||f} = 0, \mu_{f} ) = 0$, which could perhaps be relaxed by changing the ordering of $\mu_{f}$, but simplifies the calculation (as will be seen in \ref{app:elecKinEq}).

The derivation starts by considering the ion kinetic equation to show that the ion distribution function does not have a high velocity tail. Since the current drive source does not explicitly appear, $\tilde{S}^{Ip}_{e}$ can only affect ions indirectly through two quantities: the electron distribution function or the electromagnetic fields. First, the electron distribution function only appears in the ion kinetic equation through the collision operator. Since the collisionality scales asymptotically as $v^{-3}$, collisions that are ordered to be $O(1)$ in the thermal part of the distribution become much weaker (i.e. $O( \epsilon^{3})$) in the high velocity tail. Moreover, we will find that the lowest order electron distribution has no high velocity tail, making this effect even weaker. Therefore, collisions with electrons drive a negligibly small high velocity ion tail. 
Second, though the electromagnetic fields can be modified by a fast electron tail, they themselves are not functions of velocity, so they do not drive instability at high velocities. Thus, while $\tilde{S}^{Ip}_{e}$ {\it can} affect ion behavior at ion thermal speeds through the fields, it is unable to excite a high velocity tail in the ion distribution function. Therefore, the ion distribution function only has activity at thermal speeds, which is governed by the standard ion gyrokinetic equation
\begin{align}
  \frac{\partial h_{i0}}{\partial t} &+ v_{||} \hat{b} \cdot \Nabla h_{i0} + \left( \vec{c}_{\kappa} v_{||}^{2} + \vec{c}_{\nabla B} \mu \right) \cdot \Nabla h_{i0} + c_{a||} \mu \frac{\partial h_{i0}}{\partial v_{||}} - \sum_{s} \left\langle C^{L}_{i s 0} \right\rangle_{\varphi} \label{eq:GKeqIons} \\
  &- \frac{1}{B} \left( \Nabla h_{i0} \times \Nabla \langle \chi_{0} \rangle_{\varphi} \right) \cdot \hat{b} = \frac{Z_{i} e F_{Mi}}{T_{i}} \frac{\partial \langle \chi_{0} \rangle_{\varphi}}{\partial t} + \frac{1}{B} \left( \Nabla F_{Mi} \times \Nabla \langle \chi_{0} \rangle_{\varphi} \right) \cdot \hat{b} , \nonumber
\end{align}
where $\langle \chi_{0} \rangle_{\varphi} = \langle \phi_{0} \rangle_{\varphi} - v_{||} \langle A_{||0} \rangle_{\varphi} - \langle \vec{v}_{\perp} \cdot \vec{A}_{\perp0} \rangle_{\varphi}$.

Next we will consider the electron kinetic equation, which is considerably more complicated to derive as it contains the current drive source on the right side. Thus, we have relegated the details of the derivation to \ref{app:elecKinEq} and will only summarize it here. We begin by taking the drift kinetic limit $\vec{\rho}_{e} \ll \vec{x} \sim \vec{X}$ of the electron gyrokinetic equation (even for the high velocity electron tail as we have assumed that $v_{f} \ll (m_{i}/m_{e}) v_{thi}$). We use drift kinetic electrons since it is a realistic assumption that simplifies the calculation, and we are only seeking a reasonable and internally consistent model for ion-scale turbulence. However, we believe it is possible to generalize the calculation to gyrokinetic electrons by adjusting the orderings of $v_{||f}$, $\mu_{f}$, and $\tilde{S}^{Ip}_{e}$ somewhat. Then we substitute $v_{||} \rightarrow v_{||} + v_{||f}$ and $\mu \rightarrow \mu + \mu_{f}$ and perform the multi-scale analysis by expanding order by order in $\epsilon \ll 1$. Due to the magnitude of the current drive source $\tilde{S}^{Ip}_{e} ( v_{||f}, \mu_{f} ) \sim \epsilon h_{e} v_{the}/a$, it does not appear until the third order of the expansion, meaning that the zeroth, first, and second order distribution functions have no high velocity tail. Thus, we find that the lowest order electron distribution function only has a contribution at thermal velocity scales and is governed by equation \refEq{eq:DKeqElec0vth}, which is identical {\it in form} to the standard drift kinetic equation. However, as with the ion kinetic equation (i.e. equation \refEq{eq:GKeqIons}), this doesn't imply that the current source has no effect. Specifically, the electromagnetic fields can still be modified by the current source through the field equations and affect the lowest order electron behavior at thermal velocity scales. To see if this is the case, we must continue in our expansion in $\epsilon$ to find the lowest order effect of the source $\tilde{S}^{Ip}_{e}$. At third order, it appears and balances against the curvature drift to determine the lowest order fast electron distribution function
\begin{align}
\frac{\partial h_{e3}^{f}}{\partial r} = \frac{\tilde{S}^{Ip}_{e0} ( v_{||f}, \mu_{f} )}{v_{||f}^{2} \vec{c}_{\kappa} \cdot \Nabla r} , \label{eq:he3vf}
\end{align}
where the superscript $f$ in $h_{e3}^{f}$ is simply a reminder that it contains a contribution at fast velocity scales. Here we've adopted the standard field-aligned coordinate system $( \alpha, r, \theta)$ typical of local gyrokinetic codes \cite{BeerBallooningCoordinates1995}, where %we've chosen to replace $r$ with the poloidal magnetic flux $\psi$,
\begin{align}
  \alpha ( r, \theta, \zeta ) \equiv \zeta - q (r) \theta \label{eq:alphaDef}
\end{align}
is the binormal spatial coordinate, $\theta$ is the straight-field line poloidal angle, and $\zeta$ is the toroidal angle.

To understand the impact of this fast electron tail, we must also consider the electromagnetic fields, which are calculated through the quasineutrality equation and Ampere's law. We will give a summary of the derivation here, while the details can be found in \ref{app:fieldEqs}. As done above, we start by taking the drift kinetic limit for electrons, this time in equations \refEq{eq:fieldPhi} through \refEq{eq:fieldBpar}. Then we expand to lowest order in $\epsilon \ll 1$, finding that the fast electron tail is one order too small to contribute to the lowest order quasineutrality equation. This means that the charge density and, hence, the electrostatic potential are determined solely by the distribution functions at thermal velocity scales. However, in Ampere's law the fast electron tail is one order larger (because the electric current is proportional to velocity), so it is the same size as the thermal contribution. Thus, the current source {\it does} affect the dynamics at thermal velocity scales --- it drives a high velocity tail in the electron distribution function (i.e. equation \refEq{eq:he3vf}), which modifies the electric current in Ampere's law and alters the lowest order perturbed magnetic field through $\delta B_{|| 0}$ and $A_{|| 0}$. Because Ampere's law is linear in both $A_{||}$ and $\delta B_{||}$, we can choose to divide the perturbed magnetic field into the portion arising from the thermal distribution and a new portion from the fast electron tail according to $A_{|| 0} = A_{|| 0}^{th} + A_{|| 0}^{f}$ and $\delta B_{|| 0} = \delta B_{|| 0}^{th} + \delta B_{|| 0}^{f}$. The thermal component of these fields must satisfy the standard Ampere's law already solved by gyrokinetic codes (i.e. equations \refEq{eq:fieldAparDriftExpThermal} and \refEq{eq:fieldBparDriftExpThermal}), while the fast component of these fields must satisfy
\begin{align}
  - \nabla_{\perp}^{2} A_{|| 0}^{f} &= \mu_{0} \left( 2 \pi Z_{e} e B \int_{-\infty}^{\infty} dv_{||f} \int_{0}^{\infty} d \mu_{f} ~ v_{||f} h_{e3}^{f} \right) \label{eq:fieldAparDriftExpFast} \\
  \Nabla \delta B_{|| 0}^{f} \times \hat{b} &= \mu_{0} \left( - 2 \pi m_{e} B \int_{-\infty}^{\infty} dv_{||f} \int_{0}^{\infty} d \mu_{f} ~ \mu_{f} \Nabla h_{e3}^{f} \times \hat{b} \right) . \label{eq:fieldBparDriftExpFast}
\end{align}

From equations \refEq{eq:he3vf}, \refEq{eq:fieldAparDriftExpFast}, and \refEq{eq:fieldBparDriftExpFast} we see that, since $\tilde{S}^{Ip}_{e}$ is independent of $t$ and $\alpha$, so are $A_{|| 0}^{f}$ and $\delta B_{|| 0}^{f}$. Thus, in the kinetic equations for electrons and ions, $A_{|| 0}^{f}$ and $\delta B_{|| 0}^{f}$ are eliminated by the time derivative and drop out of the turbulent drive term too. They only survive through the nonlinear term. Substituting $A_{|| 0} = A_{|| 0}^{th} + A_{|| 0}^{f}$, $\delta B_{|| 0} = \delta B_{|| 0}^{th} + \delta B_{|| 0}^{f}$, and the form of $\vec{c}_{\nabla B}$ into the standard electron drift kinetic equation (i.e. equation \refEq{eq:DKeqElec0vth}), we see that the electron kinetic equation can be written as
\begin{align}
  \frac{\partial h_{e0}^{th}}{\partial t} &+ v_{||} \left( \hat{b} + \frac{1}{B} \frac{\partial A_{|| 0}^{f}}{\partial r} \Nabla r \times \hat{b} \right) \cdot \Nabla h_{e0}^{th} + \vec{c}_{\kappa} v_{||}^{2} \cdot \Nabla h_{e0}^{th} + \frac{\mu}{\Omega_{e}} \left( \hat{b} \times \Nabla \left( B + \delta B_{|| 0}^{f} \right) \right) \cdot \Nabla h_{e0}^{th} \nonumber \\
  &+ c_{a||} \mu \frac{\partial h_{e0}^{th}}{\partial v_{||}} - \sum_{s} C^{L}_{e s 0} - \frac{1}{B} \left( \Nabla h_{e0}^{th} \times \Nabla \chi_{0}^{th} \right) \cdot \hat{b} \label{eq:finalElecEq} \\
  &= \frac{Z_{e} e F_{Me}}{T_{e}} \frac{\partial \chi_{0}^{th}}{\partial t} + \frac{1}{B} \left( \Nabla F_{Me} \times \Nabla \chi_{0}^{th} \right) \cdot \hat{b} , \nonumber
\end{align}
where $\chi_{0}^{th} = \phi_{0} - v_{||} A_{|| 0}^{th} + m_{e} \mu/ ( Z_{e} e ) \delta B_{|| 0}^{th}$. This equation is identical to the standard electron drift kinetic equation except for two terms: the one proportional to $\partial A_{|| 0}^{f}/\partial r$ and the one proportional to $\delta B_{|| 0}^{f}$. These arise from the current drive source $\tilde{S}^{Ip}_{e}$ creating a high velocity electron tail that modifies the magnetic field. The perpendicular magnetic field modification $A_{|| 0}^{f}$ alters the parallel streaming term as particles follow the modified magnetic field lines, rather than the original background field. The modification to the parallel component of the magnetic field $\delta B_{|| 0}^{f}$ changes the local field strength, thereby altering the grad-B drift. Similarly, we can separate the fast and thermal components of the fields in equation \refEq{eq:GKeqIons} and write the ion kinetic equation as 
\begin{align}
  \frac{\partial h_{i0}}{\partial t} &+ v_{||} \left( \hat{b} + \frac{1}{B} \frac{\partial \langle A_{|| 0}^{f} \rangle_{\varphi}}{\partial r} \Nabla r \times \hat{b} \right) \cdot \Nabla h_{i0} + \vec{c}_{\kappa} v_{||}^{2} \cdot \Nabla h_{i0} \label{eq:finalIonEq} \\
  &+ \frac{\mu}{\Omega_{i}} \left( \hat{b} \times \Nabla \left( B - \frac{\Omega_{i}}{\mu B} \left\langle \vec{v}_{\perp} \cdot \vec{A}_{\perp0}^{f} \right\rangle_{\varphi} \right) \right) \cdot \Nabla h_{i0} + c_{a||} \mu \frac{\partial h_{i0}}{\partial v_{||}} - \sum_{s} \langle C^{L}_{i s} \rangle_{\varphi} \nonumber \\
  &- \frac{1}{B} \left( \Nabla h_{i0} \times \Nabla \langle \chi_{0} \rangle_{\varphi}^{th} \right) \cdot \hat{b} = \frac{Z_{i} e F_{Mi}}{T_{i}} \frac{\partial \langle \chi_{0} \rangle_{\varphi}^{th}}{\partial t} + \frac{1}{B} \left( \Nabla F_{Mi} \times \Nabla \langle \chi_{0} \rangle_{\varphi}^{th} \right) \cdot \hat{b} , \nonumber
\end{align}
where $\langle \chi_{0} \rangle_{\varphi}^{th} \equiv \langle \phi_{0} \rangle_{\varphi} - v_{||} \langle A_{|| 0}^{th} \rangle_{\varphi} - \langle \vec{v}_{\perp} \cdot \vec{A}_{\perp 0}^{th} \rangle_{\varphi}$. Thus, we see that the ion gyrokinetic equation has modifications analogous to the electron drift kinetic equation.

In the remainder of this paper, we will ignore the modification to the grad-B drift from the current source (i.e. $\delta B_{||0}^{f} = \vec{A}_{\perp0}^{f} = 0$). It is an interesting physical effect worth exploring and implementing in gyrokinetic codes, as we expect it to have just as large of an impact as the modifications to the parallel streaming term. However, it is outside the scope of the present paper. Instead we will focus on the new parallel streaming term and show how it can represent a modification to the safety factor profile. To demonstrate this, we will include an arbitrary safety factor modification $\tilde{q} (r)$ in the standard binormal coordinate $\alpha$ (defined by equation \refEq{eq:alphaDef}) to produce
\begin{align}
  \tilde{\alpha} &\equiv \zeta - ( q (r) + \tilde{q} (r) ) \theta \label{eq:alphaTildeDef} \\
  &= \alpha - \tilde{q} (r) \theta  . \label{eq:alphaTildeVsAlpha}
\end{align}
For simplicity we've chosen this form so that the modified field lines remain straight in the $( \theta, \zeta)$ plane. In the $(\tilde{\alpha}, r, \theta)$ coordinate system, the ion parallel streaming term in equation \refEq{eq:finalIonEq} becomes
\begin{align}
  v_{||} \left( \hat{b} + \frac{1}{B} \frac{\partial \langle A_{|| 0}^{f} \rangle_{\varphi}}{\partial r} \Nabla r \times \hat{b} \right) \cdot \Nabla h_{i0} &= v_{||} \hat{b} \cdot \Nabla \theta \left. \frac{\partial h_{i0}}{\partial \theta} \right|_{\tilde{\alpha}} \label{eq:parStreamTilde} \\
  &+ v_{||} \left( \frac{\partial \langle A_{|| 0}^{f} \rangle_{\varphi}}{\partial \psi} - \hat{b} \cdot \Nabla \theta ~ \tilde{q} (r) \right) \frac{\partial h_{i0}}{\partial \tilde{\alpha}} , \nonumber
\end{align}
after ignoring several small terms in $\rho_{\ast} \ll 1$. Note that we've replaced $r$ in some places with the poloidal magnetic flux $\psi$ because $\vec{B} = \Nabla \alpha \times \Nabla \psi$. Thus, we see that $( \tilde{\alpha}, r, \theta )$ corresponds to an exactly field-aligned coordinate system if
\begin{align}
  \tilde{q} (r)= \frac{1}{\hat{b} \cdot \Nabla \theta} \frac{\partial \langle A_{|| 0}^{f} \rangle_{\varphi}}{\partial \psi}  . \label{eq:qTildeAparRelation}
\end{align}
Combining this result with equations \refEq{eq:he3vf} and \refEq{eq:fieldAparDriftExpFast}, we find
\begin{align}
  \frac{\partial^{2} \tilde{q}}{\partial r^{2}} = - \frac{2 \pi \mu_{0} Z_{e} e B}{ (\hat{b} \cdot \Nabla \theta) ( \vec{c}_{\kappa} \cdot \Nabla \psi ) | \Nabla r |^{2} }\int_{-\infty}^{\infty} dv_{||f} \int_{0}^{\infty} d \mu_{f} ~ \frac{\langle \tilde{S}^{Ip}_{e0} \rangle_{\varphi} ( v_{||f}, \mu_{f} )}{v_{||f}} . \label{eq:qTildeSIpRelation}
\end{align}
It is not necessarily possible to find a $\tilde{q}(r)$ that satisfies this equation, except for particular choices of $\tilde{S}^{Ip}_{e0}$. This is because the left side of the equation depends only on minor radius, while the right side can also depend on species and magnetic moment (through the gyroaverage) and poloidal angle (through the geometric factors and $\tilde{S}^{Ip}_{e0}$). However, we remind the reader that the aim of this work is to model variation with a scale that is tens to hundreds of ion gyroradii. Thus, if we choose $\tilde{S}^{Ip}_{e0}$ to vary radially over distances significantly longer than the ion gyroradius (i.e. $\partial \tilde{S}^{Ip}_{e0} / \partial r \ll \tilde{S}^{Ip}_{e0} / \rho_{i}$), then the gyroaverage vanishes (i.e. $\langle \tilde{S}^{Ip}_{e0} \rangle_{\varphi} = \tilde{S}^{Ip}_{e0}$ as well as $\langle A_{|| 0}^{f} \rangle_{\varphi} = A_{|| 0}^{f}$) and the species and magnetic moment dependences with it. Similarly, we are free to choose
\begin{align}
  \tilde{S}^{Ip}_{e0} \propto \frac{( \hat{b} \cdot \Nabla \theta ) (\vec{c}_{\kappa} \cdot \Nabla \psi) | \Nabla r |^{2}}{B} \label{eq:sourcePolVariation}
\end{align}
so that the right side of equation \refEq{eq:qTildeSIpRelation} is independent of $\theta$ (with the caveat that the poloidal locations where $\vec{c}_{\kappa} \cdot \Nabla \psi = 0$ must be treated properly as is done in \ref{app:radialDrift}). Thus, equation \refEq{eq:qTildeSIpRelation} shows that, by carefully choosing $\tilde{S}^{Ip}_{e0}$, we can create {\it any} radially-periodic safety factor perturbation, so long as $\tilde{q} (r)$ is long wavelength compared to the ion gyroradius. {\it Ultimately, this means we can specify $\tilde{q}(r)$ as a free function instead of having to specify $\tilde{S}^{Ip}_{e0}$.}

Note that if you wanted to study a shorter wavelength current modification, you can by specifying $A_{|| 0}^{f} (r)$ instead of $\tilde{q}(r)$ (while also retaining the ion gyroaverage). Additionally, we chose the form of equation \refEq{eq:alphaTildeDef} for simplicity, but other choices with a more complicated dependence on $\theta$ are also possible. These would be represented by a $\tilde{q} (r, \theta)$ that depends on $\theta$ and would correspond to different functional forms of $\tilde{S}^{Ip}_{e0}$ through equation \refEq{eq:qTildeSIpRelation}. Adding $\theta$ variation to $\tilde{q}$ would have the effect of altering the {\it local} magnetic shear, in addition to the global shear, which could be interesting to explore. If you wanted to model a given ECCD source as faithfully as possible, you would just specify $\tilde{S}^{Ip}_{e0}$ directly and the modification to the safety factor would be an output of the calculation.

To derive a form suitable for implementation in a gyrokinetic code, we will substitute equation \refEq{eq:qTildeAparRelation} into the parallel streaming term of the kinetic equations and use the standard $(\alpha, r, \theta)$ coordinate system to find
\begin{align}
  v_{||} \left( \hat{b} + \frac{1}{B} \frac{\partial A_{|| 0}^{f}}{\partial r} \Nabla r \times \hat{b} \right) \cdot \Nabla h_{s0} &= v_{||} \hat{b} \cdot \Nabla \theta \left( \left. \frac{\partial h_{s0}}{\partial \theta} \right|_{\alpha} + \tilde{q} (r) \frac{\partial h_{s0}}{\partial \alpha} \right) \label{eq:parStream} .
\end{align}
We've chosen to use $\alpha$ instead of $\tilde{\alpha}$ because geometric coefficients that appear elsewhere in the gyrokinetic equation (e.g. $\Nabla r \cdot \Nabla \tilde{\alpha}$) would gain an explicit dependence on $r$ from the ion-scale variation of $\tilde{q} (r)$, complicating the Fourier space treatment typically used. Additionally, keeping the standard coordinates allows us to maintain all of the same flux tube boundary conditions (including the twist-and-shift parallel condition \cite{BeerBallooningCoordinates1995}) without needing any modifications. However, the downside of using $\alpha$ is that the coordinate system is not exactly field-aligned when the modification to the magnetic geometry is included. Specifically, increasing the amplitude of $\tilde{q} (r)$ eventually requires proportionally increasing the resolution in $\theta$. This is because, as $\tilde{q}(r)$ is increased, turbulent eddies (which stretch along {\it modified} magnetic field lines) begin to angle diagonally across the $(\alpha, \theta)$ grid. As a result, the spatial scale of the variation along the rows of $\theta$ grid points (at constant $\alpha$) can become dominated by the {\it perpendicular} variation of the eddies.

To investigate this mathematically, we can imagine an idealized turbulent perturbation written in the exactly field-aligned $( \tilde{\alpha}, r, \theta)$ coordinate system as
\begin{align}
\phi ( \tilde{\alpha}, r, \theta) = \cos \left( \lambda_{\tilde{\alpha}} \tilde{\alpha} \right) \cos \left( \lambda_{\theta} \theta \right) , \label{eq:perturbFA}
\end{align}
where $\lambda_{\tilde{\alpha}}$ and $\lambda_{\theta}$ are the wavelengths of the perturbations in the $\tilde{\alpha}$ and $\theta$ directions respectively. These wavelengths represent the typical scale of variation, so one would need grid resolutions of $\Delta \tilde{\alpha} \sim \lambda_{\tilde{\alpha}}$ and $\left. \Delta \theta \right|_{\tilde{\alpha}} \sim \lambda_{\theta}$ to properly resolve the perturbation. Next, we can take equation \refEq{eq:perturbFA}, substitute equation \refEq{eq:alphaTildeVsAlpha}, and use trigonometric identities (i.e. the angle sum and product-to-sum relations) to write the perturbation in the standard $( \alpha, r, \theta)$ coordinate system as
\begin{align}
\phi ( \alpha, r, \theta) &= \cos \left( \lambda_{\tilde{\alpha}} \alpha \right) \frac{\cos \left( \left( \tilde{q} (r) \lambda_{\tilde{\alpha}} + \lambda_{\theta} \right) \theta \right) + \cos \left( \left( \tilde{q} (r) \lambda_{\tilde{\alpha}} - \lambda_{\theta} \right) \theta \right)}{2} \label{eq:perturbConv} \\
&+ \sin \left( \lambda_{\tilde{\alpha}} \alpha \right) \frac{\sin \left( \left( \tilde{q} (r) \lambda_{\tilde{\alpha}} + \lambda_{\theta} \right) \theta \right) + \sin \left( \left( \tilde{q} (r) \lambda_{\tilde{\alpha}} - \lambda_{\theta} \right) \theta \right)}{2} . \nonumber
\end{align}
Thus, we see that while the resolution in $\alpha$ can remain similar to the resolution in $\tilde{\alpha}$, the $\theta$ resolution in the standard coordinate system $\left. \Delta \theta \right|_{\alpha}$ must scale as
\begin{align}
\left. \Delta \theta \right|_{\alpha} \sim \tilde{q} (r) \lambda_{\tilde{\alpha}} + \lambda_{\theta} \sim \tilde{q} (r) \Delta \tilde{\alpha} + \left. \Delta \theta \right|_{\tilde{\alpha}} . \label{eq:nonFieldAligned}
\end{align}
For small $\tilde{q} (r)$, the second term dominates and the required number of $\theta$ grid points remains unchanged between the two coordinate systems. However, as $\tilde{q} (r)$ is increased, the first term eventually dominates and the required parallel resolution $\left. \Delta \theta \right|_{\alpha} \sim \tilde{q} (r) \Delta \tilde{\alpha}$ becomes proportional to $\tilde{q}(r)$. Note that, because $\tilde{q} (r) \sim  \rho_{\ast} q_{0}$ the coordinate system always remains field-aligned to lowest order in $\rho_{\ast}$, so we never need parallel grid spacing on the scale of $\rho_{i}$. We just require progressively finer resolution on the equilibrium scale $a$.

Because of our decision to use $\alpha$, we can easily Fourier-analyze the parallel streaming term on the right side of equation \refEq{eq:parStream} (as well as the entire gyrokinetic model) to find
\begin{align}
  v_{||} \hat{b} \cdot \Nabla \theta \left( \left. \frac{\partial h_{s0}}{\partial \theta} \right|_{k_{r}, k_{\alpha}} \right. &+ i \frac{k_{\alpha}}{2} \sum_{n=1}^{\infty} \left[ \left( \tilde{q}^{C}_{n} + i \tilde{q}^{S}_{n} \right) h_{s0} \left( k_{r} + \frac{2 \pi}{L_{r}} n, k_{\alpha} \right) \right. \label{eq:parStreamFourier} \\
  &+ \left. \left. \left( \tilde{q}^{C}_{n} - i \tilde{q}^{S}_{n} \right) h_{s0} \left( k_{r} - \frac{2 \pi}{L_{r}} n, k_{\alpha} \right) \right] \right) , \nonumber
\end{align}
where $L_{r}$ is the radial width of the flux tube domain and $k_{r}$ and $k_{\alpha}$ are the radial and binormal Fourier wavenumbers respectively. For clarity we have explicitly written this in terms of the sine and cosine coefficients of the Fourier-analyzed the safety factor modification given by
\begin{align}
  \tilde{q} (r) =  \sum_{n=1}^{\infty} \left[ \tilde{q}^{C}_{n} \Cos{\frac{2 \pi n}{L_{r}} (r-r_{0})} + \tilde{q}^{S}_{n} \Sin{\frac{2 \pi n}{L_{r}} (r-r_{0})} \right] . \label{eq:qTildeForm}
\end{align}
Hereafter, for brevity we will use the coefficients of the exponential form of the Fourier series, $\tilde{q}^{E}_{n} \equiv ( \tilde{q}^{C}_{n} - i \tilde{q}^{S}_{n} )/2$ and $\tilde{q}^{E}_{-n} \equiv ( \tilde{q}^{C}_{n} + i \tilde{q}^{S}_{n} )/2$ (emphasizing that $\tilde{q}^{C}_{n} = \tilde{q}^{S}_{n} = 0$ for all $n \leq 0$). Note that we have ignored $\tilde{q}^{C}_{0}$ as it is an infinitesimal perturbation to $q_{0}$. We see from equation \refEq{eq:parStreamFourier} that turbulence beats against the various radial wavenumbers of the non-uniform safety factor profile to drive activity at other radial wavenumbers through the three-wave coupling mechanism \cite{HasegawaThreeWaveCoupling1979}. Note that this three-wave coupling persists in linear as well as nonlinear calculations. Additionally, we see that one does {\it not} need to specify the current drive source $\tilde{S}^{Ip}_{e}$ and add equations \refEq{eq:DKeqElec1} and \refEq{eq:fieldAparDriftExpFast} to the gyrokinetic model. Instead one can simply specify the Fourier coefficients of the safety factor modification and solve the standard gyrokinetic system with the changes to the parallel streaming term given by equation \refEq{eq:parStreamFourier}. Any long-wavelength choice for $\tilde{q} (r)$ corresponds to a physically possible current drive source through equation \refEq{eq:qTildeSIpRelation}.

Importantly, equation \refEq{eq:parStreamFourier} is straightforward to implement in a gyrokinetic code. This has been done for the local gyrokinetic code GENE \cite{JenkoGENE2000}. In practice, we chose to specify the Fourier coefficients of the magnetic shear profile $\tilde{s} (r) = (r_{0}/q_{0}) d\tilde{q}/dr$, $\tilde{s}^{C}_{n} =( r_{0} / q_{0} ) ( 2 \pi n / L_{r}) \tilde{q}^{S}_{n}$ and $\tilde{s}^{S}_{n} = - ( r_{0} / q_{0} ) ( 2 \pi n / L_{r}) \tilde{q}^{C}_{n}$, because the magnitude of these coefficients can be directly compared against the background magnetic shear $\hat{s}$ value (e.g. setting $\tilde{s}^{C}_{1} = \hat{s}$ and all other Fourier coefficients to zero will exactly cancel the effective magnetic shear at $r=r_{0}$ regardless of the radial box size $L_{r}$).

%%===================================================%
%%===================================================%
\section{Benchmarking}
\label{sec:benchmark}
%%===================================================%
%%===================================================%

We will perform two benchmarks, one linear and one nonlinear, to verify our computational implementation of non-uniform magnetic shear in GENE.

In the first benchmark, we will use the cold ion limit and an adiabatic treatment of electrons to enable the analytic calculation of linear growth rates in the presence of non-uniform magnetic shear. Though not the most realistic, these simplifications are often used together to facilitate the study of plasma dynamics \cite{CattoPVG1973, HasegawaMimaEquation1978, ZhuColdIonLimitGyro2010, JorgeColdIonGBS2017, IvanovDimitsRegime2020}. We will start from the derivation in \cite{BallFlowShear2019}, which models Parallel Velocity Gradient (PVG) turbulence in a slab geometry. By assuming the cold ion limit $T_{i} \ll Z_{i} T_{e}$ and adiabatic electrons, it rigorously derives two coupled fluid equations that exactly correspond to the full electrostatic, collisionless gyrokinetic model. No fluid closure is required. We will start from equations (16) and (17) in \cite{BallFlowShear2019}, but include the modifications to the parallel streaming term arising from non-uniform magnetic shear (i.e. equation \refEq{eq:parStreamFourier}). The density evolution equation (after enforcing quasineutrality) is given by 
\begin{align}
  &\left( 1 + k_{x}^{2} \rho_{S}^{2} + k_{y}^{2} \rho_{S}^{2} \right) \frac{\partial \phi}{\partial t} \label{eq:densityMom} \\
  &\hspace{3em} + \frac{T_{e}}{e} \left( \frac{\partial \delta u_{||}}{\partial z} + i k_{y} \sum_{n=1}^{\infty} \left[ \tilde{q}^{E}_{-n} \delta u_{||} \left( k_{x} + \frac{2 \pi}{L_{x}} n \right) + \tilde{q}^{E}_{n} \delta u_{||} \left( k_{x} - \frac{2 \pi}{L_{x}} n \right) \right] \right) = 0 \nonumber
\end{align}
and the parallel velocity evolution equation is
\begin{align}
\frac{\partial \delta u_{||}}{\partial t} &+ \frac{Z_{i} e}{m_{i}} \left( \frac{\partial \phi}{\partial z} + i k_{y} \sum_{n=1}^{\infty} \left[ \tilde{q}^{E}_{-n} \phi \left( k_{x} + \frac{2 \pi}{L_{x}} n \right) + \tilde{q}^{E}_{n} \phi \left( k_{x} - \frac{2 \pi}{L_{x}} n \right) \right] \right) = i \frac{k_{y}}{B} \omega_{V||} \phi , \label{eq:parVelMom}
\end{align}
where $\rho_{S} \equiv c_{S} / \Omega_{i}$ is the sound gyroradius, $c_{S} \equiv \sqrt{Z_{i} T_{e}/m_{i}}$ is the sound speed, $\omega_{V||} \equiv - d u_{||} / dx$ is the radial gradient of the background parallel flow velocity $u_{||}$, $\delta u_{||} \equiv (2 \pi B / n_{i}) \int_{-\infty}^{\infty} dv_{||} \int_{0}^{\infty} d \mu ~ v_{||} g_{i}$ is a perturbed parallel flow velocity, $g_{i} = h_{i} - (Z_{i} e F_{Mi}/T_{i}) \langle \phi \rangle_{\varphi}$ is the so-called complementary distribution function evolved by GENE, and we take the normalized coordinate system used by GENE where $x$ is analogous to $r$, $y$ to $\alpha$, and $z$ to $\theta$. To produce these two equations, we have made several choices to simplify the problem as much as possible, while still appropriately testing the computational implementation of the non-uniform magnetic shear. Specifically, we have assumed the most basic slab geometry in order to neglect the magnetic drifts and simplify all geometric factors ($\omega_{Mx} = \omega_{My} = \partial B / \partial z = 0$ and $|\Nabla x| = |\Nabla y| = |\Nabla z| = \hat{b} \cdot \Nabla z = J = 1$ in the terminology of \cite{BallFlowShear2019}). We also chose to omit the density gradients, temperature gradients, perpendicular velocity shear, and background global magnetic shear ($\omega_{V \perp} = u_{f} = \hat{s} = 0$ again in the terminology of \cite{BallFlowShear2019}). Lastly, we have ignored the nonlinear term as non-uniform magnetic shear already modifies the linear dynamics, which are much less computationally expensive to calculate with GENE.

Combining the density and parallel velocity moments of equations \refEq{eq:densityMom} and \refEq{eq:parVelMom}, then Fourier-analyzing in $\partial/\partial t \rightarrow i \omega$ and $\partial/\partial z \rightarrow i k_{z}$ gives
\begin{align}
  &\left[ \left( 1 + k_{x}^{2} \rho_{S}^{2} + k_{y}^{2} \rho_{S}^{2} \right) \omega^{2} + k_{y} \rho_{S} k_{z} c_{S} \omega_{V||} - k_{z}^{2} c_{S}^{2} \right] \phi \nonumber \\
  &\hspace{3em} + \left( k_{y}^{2} \rho_{S}^{2} \omega_{V||} - 2 k_{y} \rho_{S} k_{z} c_{S} \right) \Omega_{i} \sum_{n=1}^{\infty} \left[ \tilde{q}^{E}_{-n} \phi \left( k_{x} + \frac{2 \pi}{L_{x}} n \right) + \tilde{q}^{E}_{n} \phi \left( k_{x} - \frac{2 \pi}{L_{x}} n \right) \right] \label{eq:coldIonEvol} \\
  &\hspace{3em} - k_{y}^{2} \rho_{S}^{2} \Omega_{i}^{2} \sum_{n=1}^{\infty} \sum_{m=1}^{\infty} \left[ \tilde{q}^{E}_{-n} \tilde{q}^{E}_{-m} \phi \left( k_{x} + \frac{2 \pi}{L_{x}} ( n + m ) \right) + \tilde{q}^{E}_{-n} \tilde{q}^{E}_{m} \phi \left( k_{x} + \frac{2 \pi}{L_{x}} ( n - m ) \right) \right. \nonumber \\
  &\hspace{6em} + \left. \tilde{q}^{E}_{n} \tilde{q}^{E}_{-m} \phi \left( k_{x} + \frac{2 \pi}{L_{x}} ( -n + m ) \right) + \tilde{q}^{E}_{n} \tilde{q}^{E}_{m} \phi \left( k_{x} + \frac{2 \pi}{L_{x}} ( -n - m ) \right) \right] = 0 . \nonumber
\end{align}
Note that, when the non-uniform shear and finite sound gyroradius effects are ignored, this reduces to the typical PVG stability limit \cite{CattoPVG1973,NewtonFlowShearUnderstanding2010}. For comparison against GENE, we are interested in solving equation \refEq{eq:coldIonEvol} for $\omega$, given a single $k_{y}$ mode and a finite grid of $k_{x}$ modes. Thus, equation \refEq{eq:coldIonEvol} can be cast as an eigenvalue problem $\tensor{A} \cdot \vec{\phi}_{kx} = \lambda \vec{\phi}_{kx}$ with eigenvalues of $\lambda = 0$, where the elements of the vector $\vec{\phi}_{kx}$ are the amplitudes of $\phi$ for the different discrete $k_{x}$ modes present in the simulation. The elements of the matrix $\tensor{A}$ are given by
\begin{align}
   A_{jk} &= \left[ \left( 1 + k_{x,j}^{2} \rho_{S}^{2} + k_{y}^{2} \rho_{S}^{2} \right) ( \omega L_{z}/c_{S} )^{2} + k_{y} \rho_{S} k_{z} L_{z} ( \omega_{V||} L_{z}/c_{S} ) - k_{z}^{2} L_{z}^{2} \right] \delta_{jk} \nonumber \\
   &+ 2 \left[ k_{y}^{2} \rho_{S}^{2} ( \omega_{V||}  L_{z}/c_{S} ) - 2 k_{y} \rho_{S} k_{z} L_{z} \right] \tilde{q}^{E}_{j-k} \rho_{\ast}^{-1} \label{eq:coldIonMatrix} \\
   &- k_{y}^{2} \rho_{S}^{2} \left[ \sum_{m=j_{min}}^{j_{max}} \tilde{q}^{E}_{j-k+m} \tilde{q}^{E}_{-m} \rho_{\ast}^{-2} + \sum_{m=j_{min}}^{j_{max}+j} \tilde{q}^{E}_{j-k-m} \tilde{q}^{E}_{m} \rho_{\ast}^{-2} \right. \nonumber \\
   &\hspace{2.7em} + \left. \sum_{m=j_{min}}^{j_{max}-j} \tilde{q}^{E}_{j-k+m} \tilde{q}^{E}_{-m} \rho_{\ast}^{-2} + \sum_{m=j_{min}}^{j_{max}} \tilde{q}^{E}_{j-k-m} \tilde{q}^{E}_{m} \rho_{\ast}^{-2} \right] , \nonumber
\end{align}
where $\delta_{jk}$ is the Kronecker delta, in the cold ion limit $\rho_{\ast} = \rho_{S}/L_{z}$, $L_{z}$ is the domain length of the slab geometry, the radial modes present in the system are $k_{x,j} = k_{x,0} + 2 \pi j / L_{x}$ for $j \in [ j_{min}, j_{max} ]$, and the upper bounds of the summations are carefully chosen to prevent modes from coupling to parallel velocity perturbations $\delta u_{||}$ that are off the numerical grid. Since the right side of equation \refEq{eq:coldIonEvol} is zero, we are seeking the values of $\omega$ that correspond to eigenvalues of $\lambda = 0$. Thus, all the non-trivial solutions can be found by simply requiring the determinant of $\tensor{A}$ to be equal to zero. This produces a polynomial in $\omega^{2}$ with an order equal to the total number of radial modes in the grid. Thus, a closed analytic solution exists for a grid with three radial modes (i.e. the general solution to the cubic equation) and larger radial grids are straightforward to solve numerically.

\begin{figure}
	(a) \hspace{17em} (b)
	
	\begin{center}
		\includegraphics[width=0.49\textwidth]{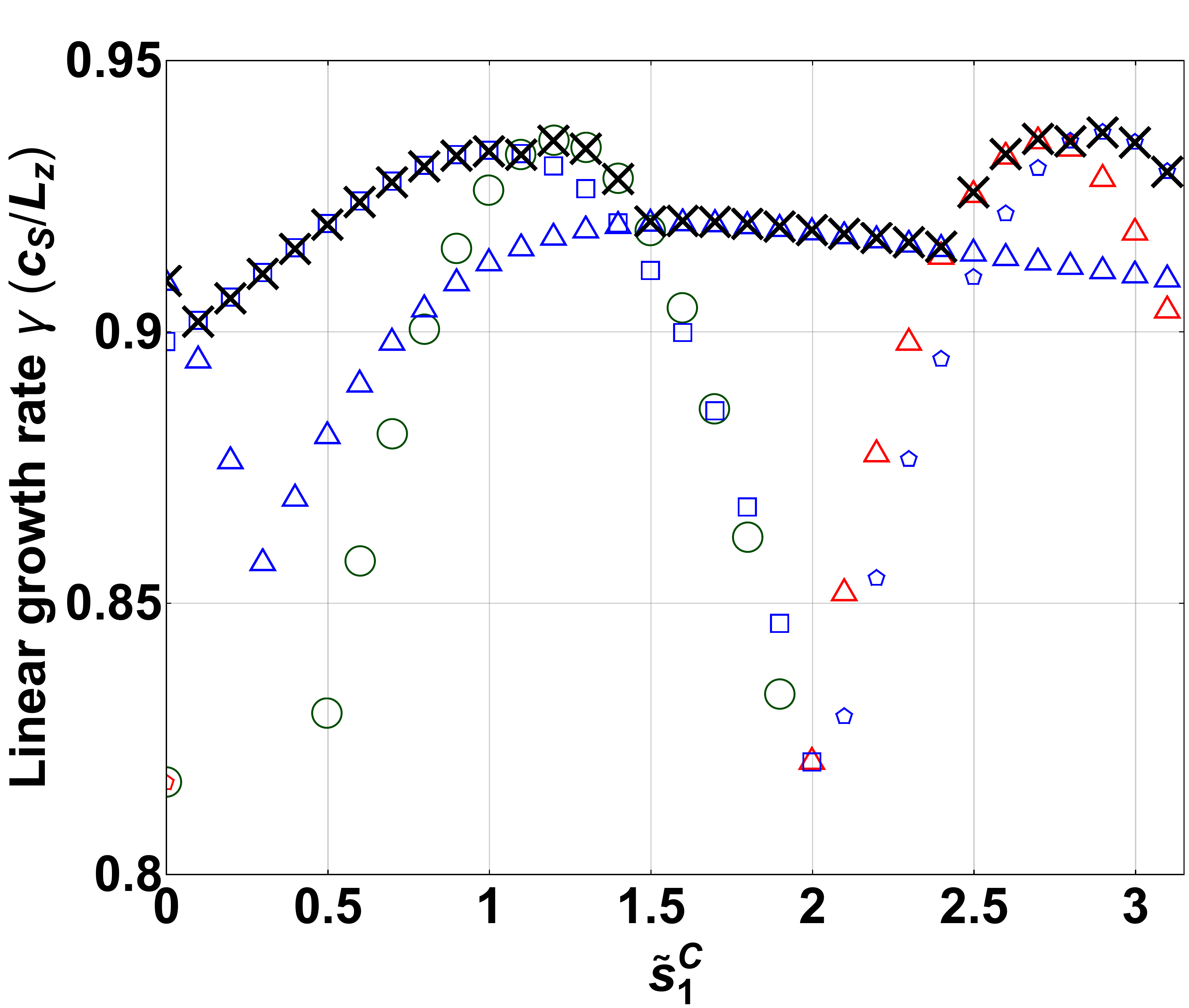}
		\includegraphics[width=0.49\textwidth]{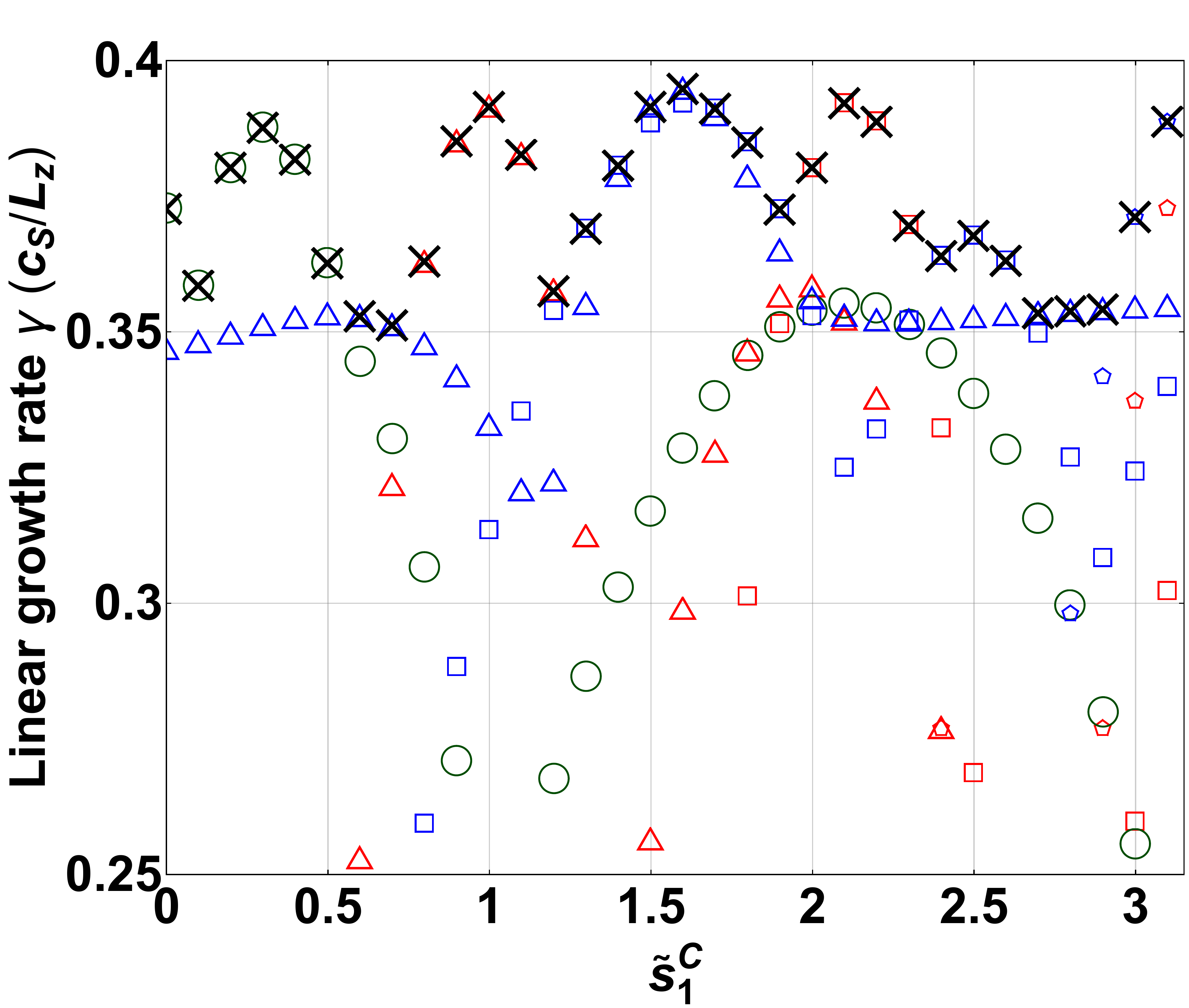}
	\end{center}
	\caption{A linear PVG benchmark in a simple slab geometry for a range of different $\tilde{s}^{C}_{1}$ values with either (a) $j_{min}=-1$, $j_{max}=1$, $\omega_{V||} L_{z} / c_{S} = 20/3$, $\tilde{s}^{S}_{1} = 0.8$, $\tilde{s}^{C}_{2} = 1$, $\tilde{s}^{S}_{2} = 1.2$ or (b) $j_{min}=-3$, $j_{max}=3$, $\omega_{V||} L_{z} / c_{S} = 3$, $\tilde{s}^{S}_{1} = 0.5$, $\tilde{s}^{C}_{2} = 0.5$, $\tilde{s}^{S}_{2} = 1.6$. The fastest-growing instabilities calculated by GENE (black crosses) are shown together with the semi-analytic result for $k_{||} L_{z} = 0$ (green circles), $k_{||} L_{z} = 1$ (blue triangles), $k_{||} L_{z} = -1$ (red triangles), $k_{||} L_{z} = 2$ (blue squares), $k_{||} L_{z} = -2$ (red squares), $k_{||} L_{z} = 3$ (blue pentagons), and $k_{||} L_{z} = -3$ (red pentagons).}
	\label{fig:linearBenchmark}
\end{figure}

Figure \ref{fig:linearBenchmark} shows a comparison between such a semi-analytic solution and GENE for two cases, which both display excellent agreement. All simulations have $T_{i}/T_{e} = 10^{-4}$, $k_{y} \rho_{S} = 0.3$, $L_{x} = 20 \rho_{S}$, and a radial grid with $k_{x,0} = 0$. All unspecified Fourier coefficients, $\tilde{s}^{S}_{n}$ and $\tilde{s}^{C}_{n}$, are zero. Since GENE discretizes the $z$ coordinate in real space, one cannot cleanly select a particular value of $k_{z}$ to simulate. Thus, to ensure that the simulation converges to the fastest-growing instability, it is important to initialize all $k_{x}$ modes using a perturbation with the fastest-growing value of $k_{z}$ (or test several different simulations initialized with different $k_{z}$ perturbations). Otherwise unstable, but sub-dominant $k_{z}$ modes can temporarily prevail causing the initial value solver of GENE to terminate prematurely and return a lower growth rate. Additionally, we note that the presence of non-uniform magnetic shear can allow PVG modes with $k_{z} = 0$ to be unstable. This initially appears surprising, given that the typical PVG instability requires a finite parallel wavenumber $k_{||}$ (i.e. a variation in $\delta u_{||}$ along the field line) \cite{CattoPVG1973, BallFlowShear2019}. However, our result is {\it not} a numerical problem --- it is a subtlety related to the definition of $k_{z}$. Since the coordinate system is no longer field-aligned due to non-uniform magnetic shear, changing the $z$ coordinate at constant $x$ and $y$ takes you across field lines. Thus, $k_{z}$ is not actually the parallel wavenumber, so, even when $k_{z} = 0$, the parallel wavenumber $k_{||}$ can be finite and enable instability.

\begin{figure}
	\hspace{3em} (a)
	
	\begin{center}
		\includegraphics[width=0.8\textwidth]{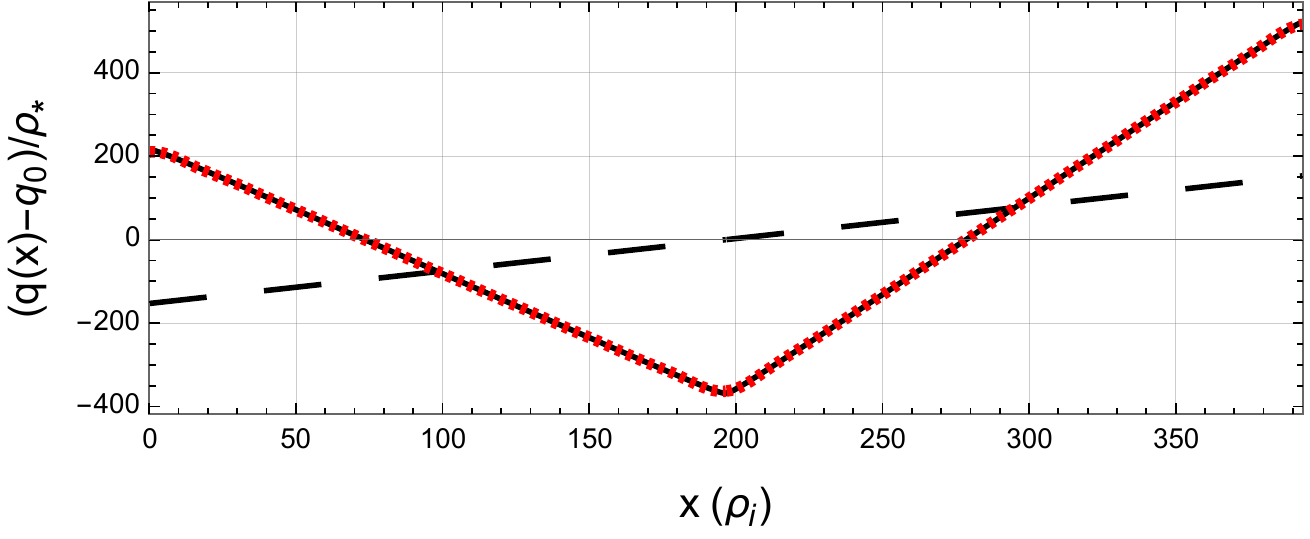}
	\end{center}
	
	\hspace{3em} (b)
	
	\begin{center}
		\includegraphics[width=0.8\textwidth]{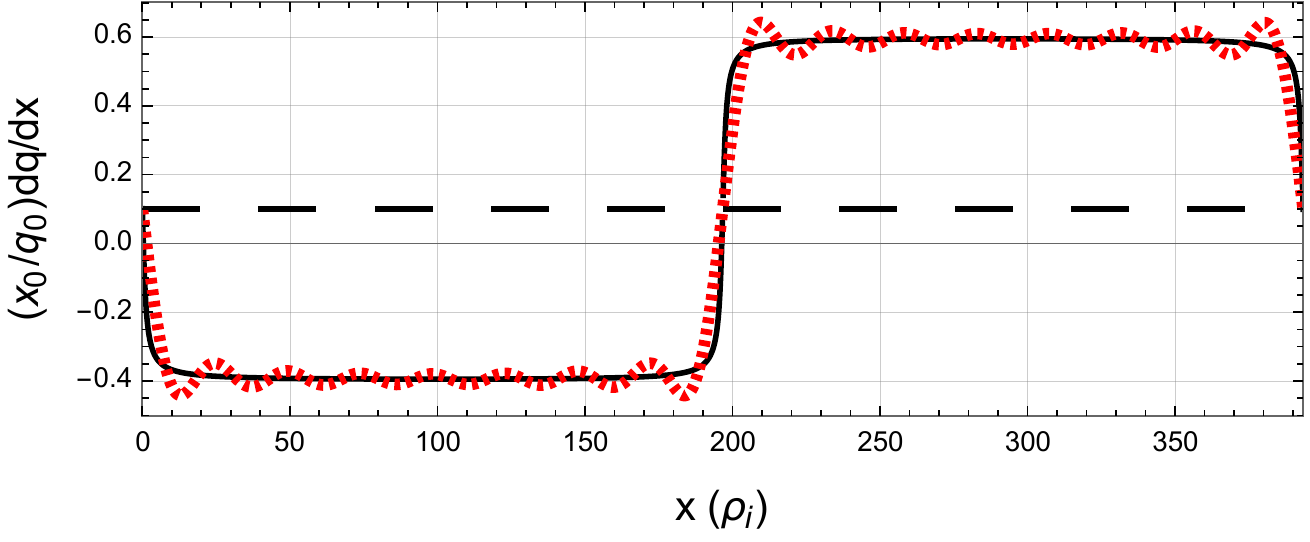}
	\end{center}
	\caption{(a) The desired total safety factor profile $q (x) + \tilde{q} (x)$ (solid black), its approximation using 16 Fourier terms (dotted red), and the unmodified safety factor profile $q(x)$ (dashed black). The solid black and dotted red profiles are nearly indistinguishable. (b) The desired total magnetic shear profile $\hat{s} + \tilde{s} (x)$ (solid black), its approximation using 16 Fourier terms (dotted red), and the unmodified magnetic shear profile $\hat{s}$ (dashed black), where $x_{0} = r_{0}/R_{0}$.}
    \label{fig:twoRegionProfiles}
\end{figure}

As a second benchmark, a nonlinear simulation was performed in tokamak geometry using parameters inspired by the Cyclone Base Case (CBC) \cite{DimitsCBC2000}. In it we employed non-uniform magnetic shear to create the safety factor profile shown in figure \ref{fig:twoRegionProfiles}. The profile has two distinct regions within a single flux tube, both with a constant value of magnetic shear. The left half has $\hat{s} = -0.4$, while the right half has $\hat{s} = 0.6$. This creates significant differences in the transport properties of the two regions, which can be compared with conventional simulations to verify our computational implementation. This ``two-region simulation'' was electrostatic, used adiabatic electrons, and ignored collisions in order to minimize computational cost. We used $16$ Fourier terms in equation \refEq{eq:parStreamFourier} to approximate the desired $\tilde{s} (x)$ profile. Adding additional terms did not substantially change the results. The other simulation parameters and resolutions are given in table \ref{tab:CBC}. Note that the cold ion limit was {\it not} used and that a large value of $N_{z}$ was chosen to ensure fully resolved turbulence despite the loss of an exactly field-aligned grid (i.e. see discussion surrounding equation \refEq{eq:nonFieldAligned}). It is also important to ensure that the simulation domain is sufficiently large in $x$ so that the dynamics are local and the middles of the two regions do not directly interact.

\begin{table}
	\centering
	\begin{tabular}{c|c||c|c}
		Parameter & Value & Parameter & Value \\
		\hline
		Minor radius of flux tube, $r_{0}/R_{0}$ & $0.18$ & Ion-e\textsuperscript{-} temperature ratio, $T_{i}/T_{e}$ & $1.0$ \\
		\hline
		Safety factor, $q_{0}$ & $1.4$ & Magnetic shear, $\hat{s}$ & $0.1$ \\
		\hline
		Temperature gradient, $R_{0}/L_{Ts}$ & $9.0$ & Density gradient, $R_{0}/L_{n}$ & $2.2$ \\
		\hline
		Ion-e\textsuperscript{-} mass ratio, $m_{i}/m_{e}$ & $3672$ & Effective ion charge, $Z_{eff}$ & $1.0$ \\
		\hline
		4\textsuperscript{th} order $\chi$ hyperdiffusion \cite{PueschelHyperDiff2010}, $\epsilon_{\chi}$ & $0.2$ & 4\textsuperscript{th} order $v_{||}$ hyperdiffusion \cite{PueschelHyperDiff2010}, $\epsilon_{v||}$ & $0.2$ \\
		\hline
		\hline
		$x / \rho_{i}$ range, $[ 0, L_{x} )$ & $[ 0, 400 )$ & Number of $x$ grid points, $N_{x}$ & $512$ \\
		\hline
		$y / \rho_{i}$ range, $[ 0, L_{y} )$ & $[ 0, 125 )$ & Number of $y$ grid points, $N_{y}$ & $64$ \\
		\hline
		$z$ range & $[ -\pi, \pi )$ & Number of $z$ points, $N_{z}$ & $128$ \\
		\hline
		$v_{||} / v_{ths}$ range & $[ - 3, 3 ]$ & Number of $v_{||}$ grid points, $N_{v||}$ & $32$ \\
		\hline
		$\sqrt{\mu / (T_{s} / B_{r})}$ range & $( 0, 3.43 ]$ & Number of $\sqrt{\mu}$ grid points, $N_{\mu}$ & $20$ \\
	\end{tabular}
	\caption{The CBC parameters \cite{DimitsCBC2000} (with a modified temperature gradient) and grid resolutions used for the two-region simulation, where the geometry is specified using the Miller model \cite{MillerGeometry1998}. Note that all grids are equally spaced and $R_{0}$ is the tokamak major radius.}
	\label{tab:CBC}
\end{table}

\begin{figure}
	\begin{center}
		\begin{tabular}{l} (a) \hspace{34em}  \end{tabular}
		
		\vspace{-0.5em}
		\includegraphics[width=\textwidth]{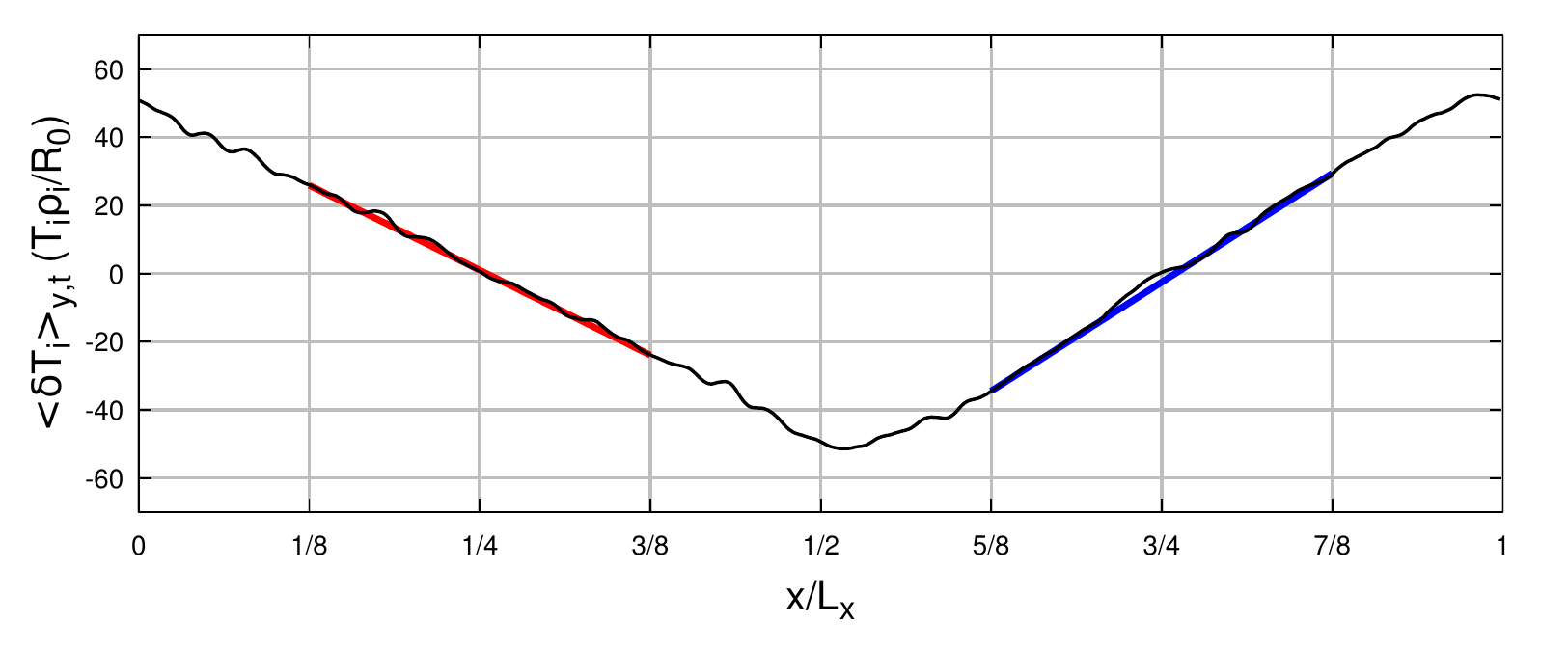}
		
		\vspace{-2em}
		\begin{tabular}{l} (b) \hspace{34em}  \end{tabular}
		
		\vspace{-0.5em}
		\includegraphics[width=\textwidth]{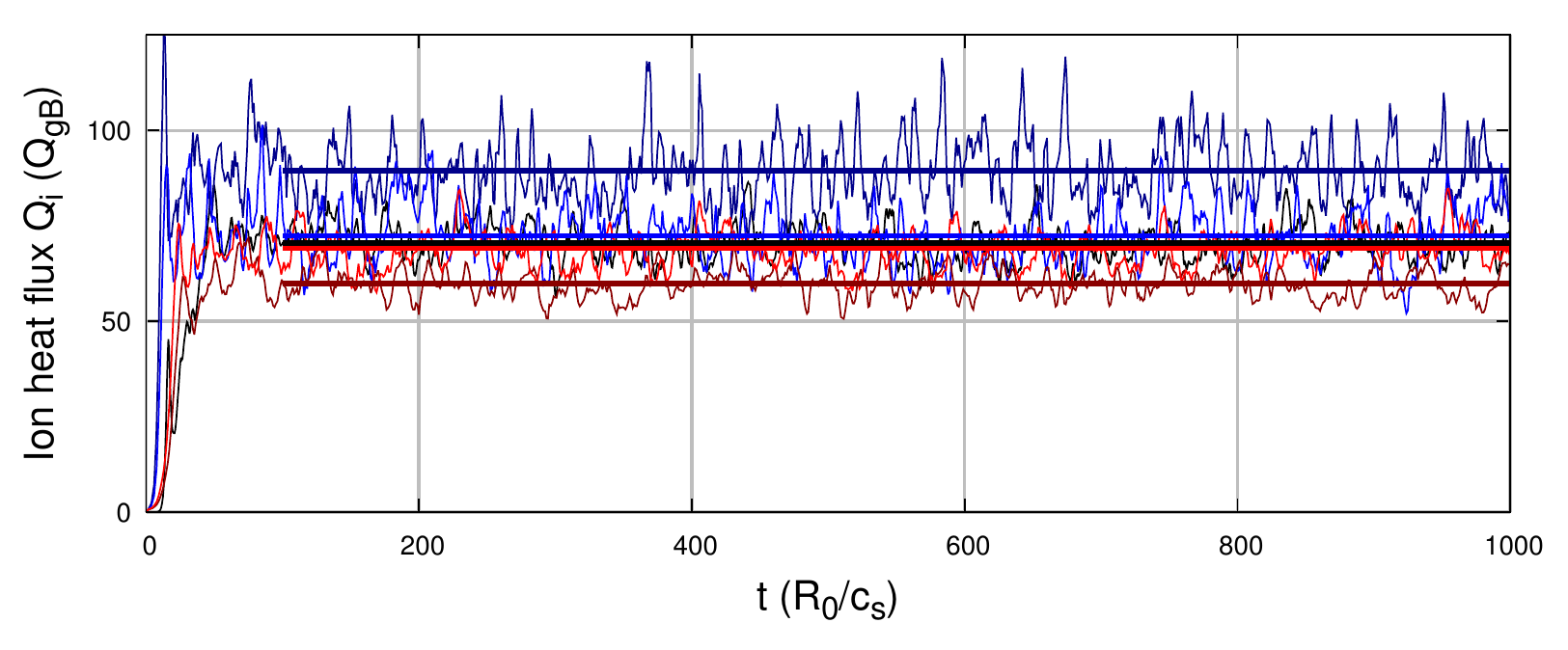}
		
		\vspace{-2em}
		\begin{tabular}{l} (c) \hspace{34em}  \end{tabular}
		
		\vspace{-0.5em}
		\includegraphics[width=\textwidth]{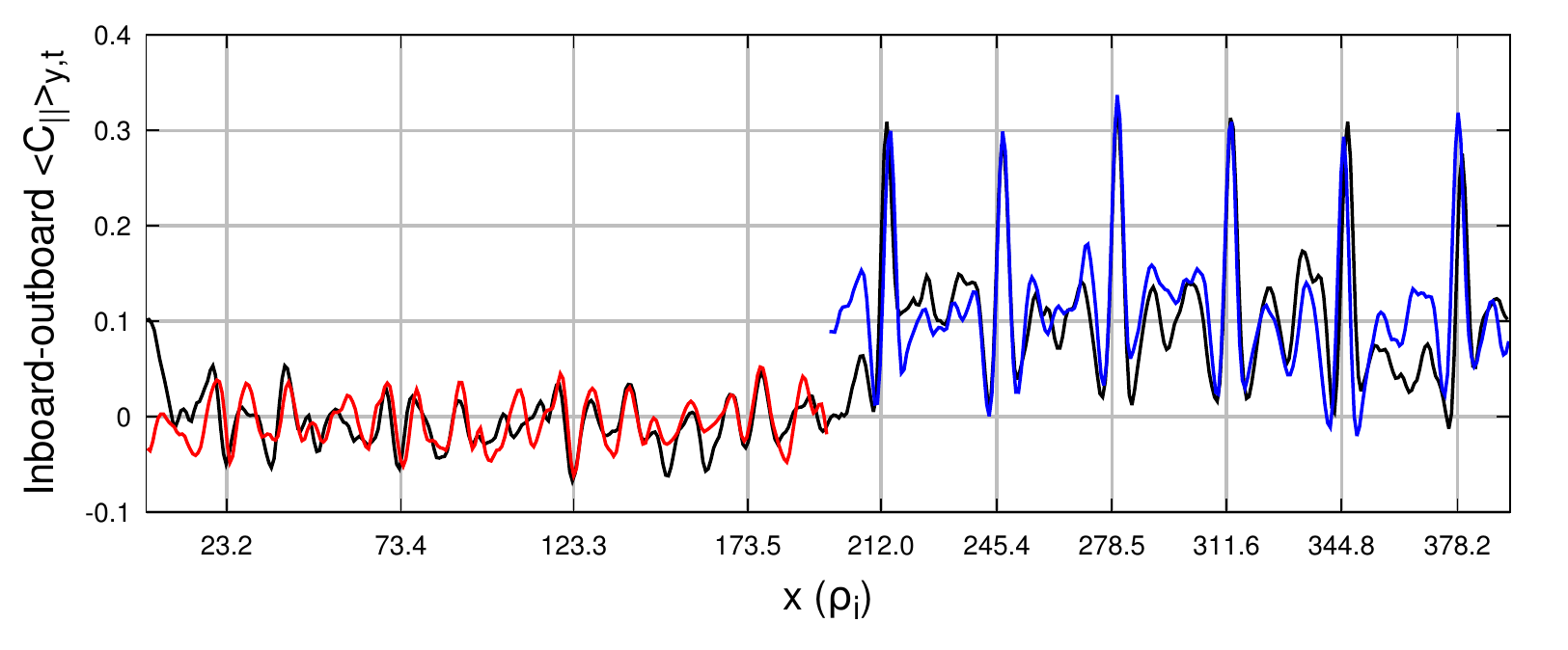}
		
		\vspace{-2em}
	\end{center}
	\caption{(a) The radial variation of the time-averaged, zonal turbulent temperature perturbation at the outboard midplane with best-fit gradients for both regions, (b) the ion heat flux time trace (thin) with its time average (thick), and (c) the two-point parallel correlation function between the inboard and outboard midplanes with the vertical grid lines indicating integer surfaces. Each plot shows the two-region, non-uniform shear simulation with $R_{0}/L_{Ti} = 9$ (black) as well as standard simulations with $\hat{s} = 0.6$ and $R_{0}/L_{Ti} = 8.3$ (blue) as well as $\hat{s} = -0.4$ and $R_{0}/L_{Ti} = 9.5$ (red). In addition, (b) shows standard simulations with $\hat{s} = 0.6$ and $R_{0}/L_{Ti} = 9$ (dark blue) as well as $\hat{s} = -0.4$ and $R_{0}/L_{Ti} = 9$ (dark red). Note that the radial domain of the standard simulations are half as wide and their correlation functions have been shifted in $x$ for ease of comparison.}
	\label{fig:twoRegionBenchmark}
\end{figure}

For the benchmark, we also ran two standard simulations to separately recreate each half of the two-region case. Thus, one simulation had $\hat{s} = -0.4$, the other had $\hat{s} = 0.6$, and neither included non-uniform magnetic shear. Both of these simulations had a radial box size and radial resolution half as large as the two-region case and could employ a lower parallel resolution of just $N_{z} = 32$. The crucial aspect of the benchmark concerns the values to take for the background gradients in the two standard simulations. Since the flux tube includes no sources of particles or energy, the quasi-steady state radial fluxes are constrained to be constant across the minor radial extent of the domain. This is true regardless of the presence of non-uniform magnetic shear. Because the local value of the magnetic shear varies across the two-region domain, the flux-gradient relationship is different at different radial locations. This means that turbulence must rearrange energy, momentum, and particles within the flux tube to create the appropriate steady-state zonal perturbations such that the fluxes are radially constant. In other words, the turbulence modifies the background gradients to ensure uniform flux profiles. Therefore, we computed the radial gradients of the time-averaged zonal temperature $\langle \delta T_{i} \rangle_{y,t}$ and density $\langle \delta n \rangle_{y,t}$ perturbations in each half of the two region domain (see figure \ref{fig:twoRegionBenchmark}(a)). Specifically, we averaged across $x/L_{x} \in [ 1/8, 3/8 )$ and $x/L_{x} \in [ 5/8, 7/8 )$ to omit the transition zones between the two regions and over $t v_{thi}/ R_{0} \in [700, 1000]$ to ensure that the zonal perturbations had time to fully form. We found the total ion temperature gradient to be $R_{0}/L_{Ti} = 9.5$ in the $\hat{s} = -0.4$ region and $R_{0}/ L_{Ti} = 8.3$ in the $\hat{s} = 0.6$ region. The direction of this result is intuitive as $\hat{s} < 0$ is usually stabilizing compared to $0 < \hat{s} < 1$ \cite{WaltzNegMagShear1995}. The density profile was not modified because the particle flux is constrained to be zero when electrons are treated adiabatically.

Figure \ref{fig:twoRegionBenchmark}(b) shows the heat flux (normalized to the gyroBohm value $Q_{gB}$) from the two-region simulation along with the standard simulations. We see from the dark red and dark blue curves that, if the temperature gradient is {\it not} adjusted, the heat fluxes disagree. However, if we use the local gradient values extracted from the respective regions of the two-region simulation (the light red and blue curves), we get good agreement. Thus, the flux-gradient relationship is the same in the standard and two-region simulations. This indicates that identical physical equilibria behave the same, regardless of whether or not they were created with non-uniform magnetic shear. Lastly, figure \ref{fig:twoRegionBenchmark}(c) shows the two-point parallel correlation between the inboard and outboard midplanes \cite{BallBoundaryCond2020}. Due to the parallel boundary condition \cite{BeerBallooningCoordinates1995}, the flux tube topology features several integer surfaces (i.e. flux surfaces with magnetic field lines that bite their own tails after one poloidal turn) \cite{BallBoundaryCond2020}. Importantly, these surfaces cause peaks in the parallel correlation and their locations can be analytically calculated from the magnetic equilibrium. In standard flux tube simulations with constant $\hat{s}$, the integer surfaces are equally spaced, but this is no longer true in the presence of non-uniform shear. For the two-region simulation, we see that the peaks occur at the calculated locations (i.e. the vertical grid lines), indicating that the non-uniform shear modifies the magnetic topology as expected. Furthermore, the widths and heights of the peaks in each half of the radial domain agree nicely with those from the corresponding standard simulation. Thus, the linear and nonlinear benchmarks give confidence that our implementation of non-uniform magnetic shear in GENE is correct.

%%===================================================%
%%===================================================%
\section{Illustrative example}
\label{sec:example}
%%===================================================%
%%===================================================%

To illustrate the possibilities that are enabled by non-uniform magnetic shear, we will use our modifications to GENE to study the importance of profile shearing in tokamaks with Positive Triangularity (PT) and Negative Triangularity (NT) plasma shaping. 

Pioneered by JET and DIII-D \cite{RebutJEThistory2018} in the 1980s, the ``D'' shaped plasma cross-section, termed positive triangularity, was found to significantly improve plasma performance. However, recent experiments on TCV \cite{CamenenTriConfinement2007, FontanaNegTriExp2017, CodaNegTriExp2021}, DIII-D \cite{AustinNegTriExp2019}, and ASDEX Upgrade \cite{HappelNegTriExp2020} have revealed that flipping the sign of triangularity to produce a negative triangularity reversed-``D'' cross-section also carries considerable advantages. While there have been relatively few gyrokinetic studies of negative triangularity \cite{MarinoniNegTri2009, MerloNegTri2015, MerloNegTri2019}, there has been one study \cite{MerloNegTriGlobal2021} specifically investigating global effects. Comparing standard nonlinear local and global GENE \cite{GoerlerGlobalGENE2011} simulations of a PT and a NT TCV \cite{CodaTCVoverview2017} equilibrium, it found that global effects were more important for NT. The two equilibria had similar safety factor profiles and total heating powers, but the background temperature profiles were considerably different (due to the different heat diffusivity in PT versus NT). Both equilibria were found to be dominated by Trapped Electron Mode (TEM) turbulence.

We will complement \cite{MerloNegTriGlobal2021} by using flux tube simulations with non-uniform magnetic shear to perform a second study of the impact of profile shearing in PT versus NT. We will base the simulations on the standard CBC \cite{DimitsCBC2000}, which is dominated by Ion Temperature Gradient (ITG) turbulence. We will include a full kinetic treatment of electrons as it was found to be important to capture the differences in transport between PT and NT. The parameters are identical to those of table \ref{tab:CBC} with a few exceptions: we use the standard magnetic shear $\hat{s} = 0.8$, strong elongation $\kappa = 1.7$, strong triangularity $\delta = \pm 0.5$, and a modified domain discretization with $L_{x} = 125 \rho_{i}$, $N_{x} = 128$, $N_{z} = 32$, $N_{v||} = 64$, and $N_{\mu} = 10$. The radial gradients of the shaping parameters were omitted for simplicity (i.e. $d \kappa / dr = d \delta / dr = 0$). The domain parameters are the result of an extensive resolution study to adequately resolve the simulation at minimal computational cost. In order to introduce profile shearing, we will add a single Fourier harmonic with $\tilde{q}_{n}^{S} / \rho_{\ast} = 35.8$ (which corresponds to $\tilde{s}_{1}^{C} = 0.25$ for $L_{x} = 125 \rho_{i}$ and $n=1$) to the safety factor profile specified according to equation \refEq{eq:qTildeForm}. Then, we will scan its wavelength $\lambda$ {\it holding the amplitude of the safety factor modification $\tilde{q}_{n}^{S}$ constant}. In practice $\lambda$ is changed using the Fourier mode number $n$ appearing in equation \refEq{eq:qTildeForm} and increasing the radial box width $L_{x}$ if needed. Note that none of the non-uniform shear simulations required an increased parallel resolution $N_{z}$ because the first term on the right side of equation \refEq{eq:nonFieldAligned} was always dominant.

\begin{figure}
	\begin{center}
		\begin{tabular}{l} (a) \hspace{29em}  \end{tabular}
		
		\vspace{-1em}
		\includegraphics[width=0.73\textwidth]{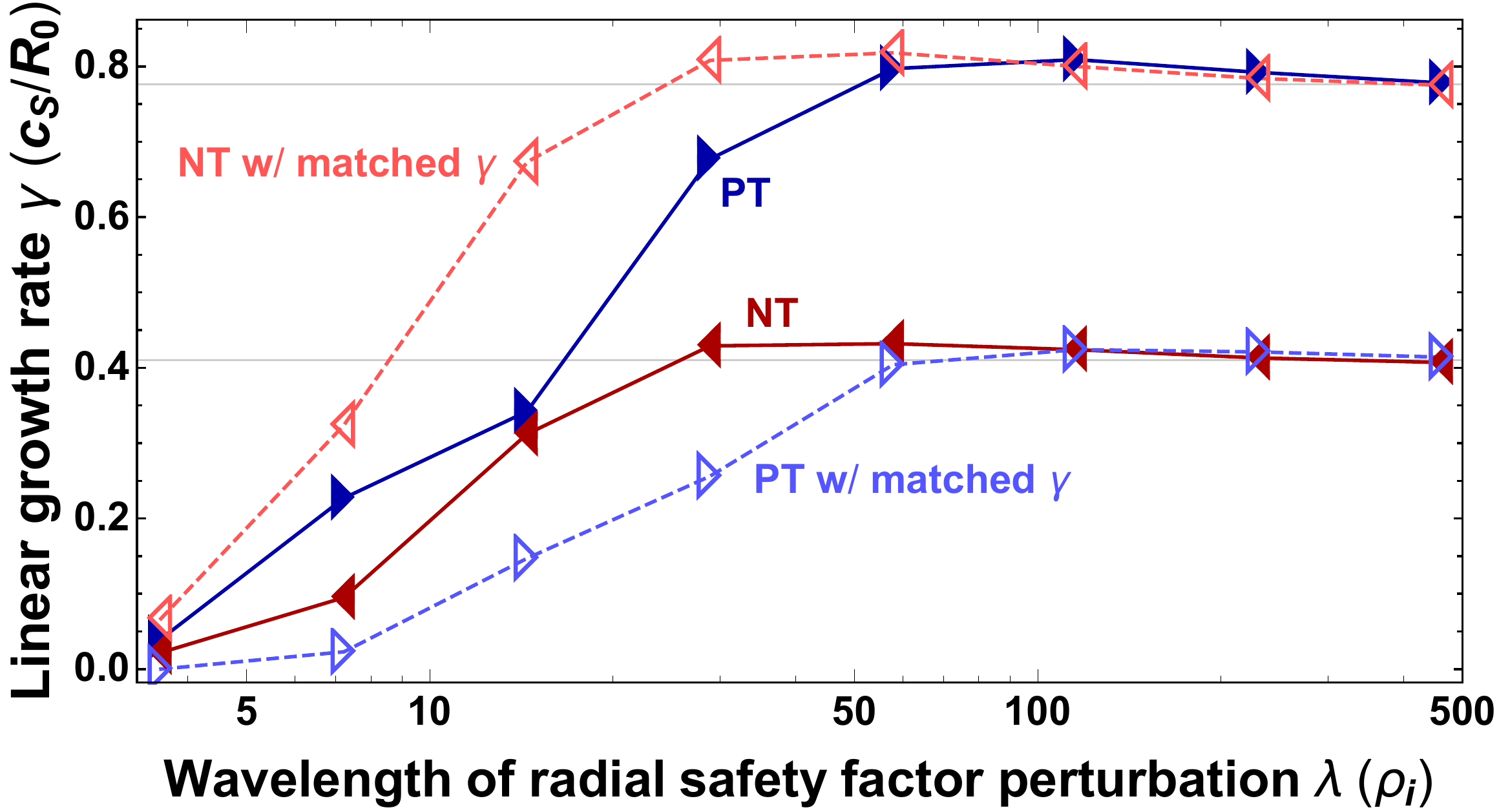}
		
		\begin{tabular}{l} (b) \hspace{29em}  \end{tabular}
		
		\vspace{-1em}
		\includegraphics[width=0.73\textwidth]{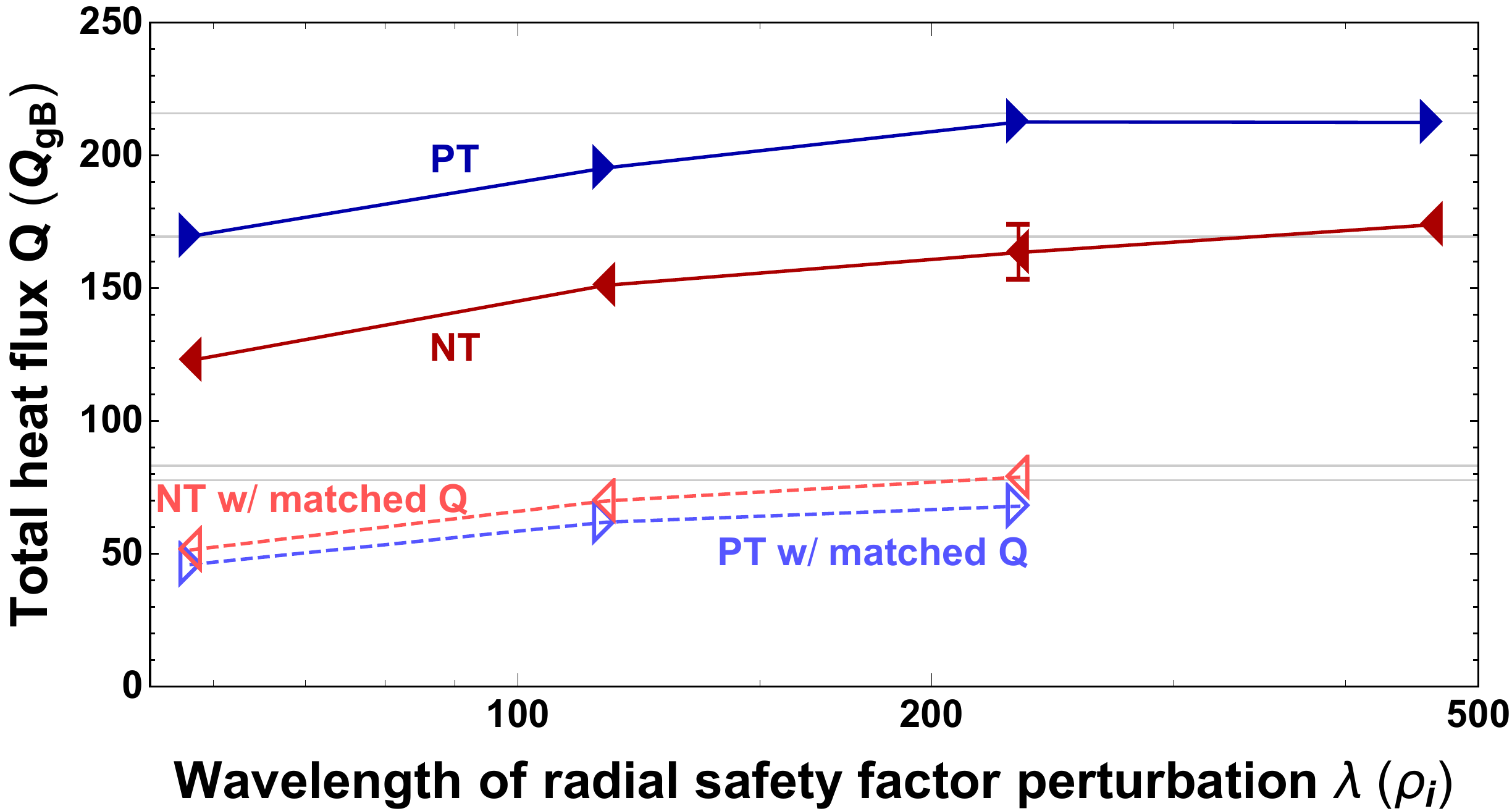}
		
		\begin{tabular}{l} (c) \hspace{29em}  \end{tabular}
		
		\vspace{-1em}
		\includegraphics[width=0.73\textwidth]{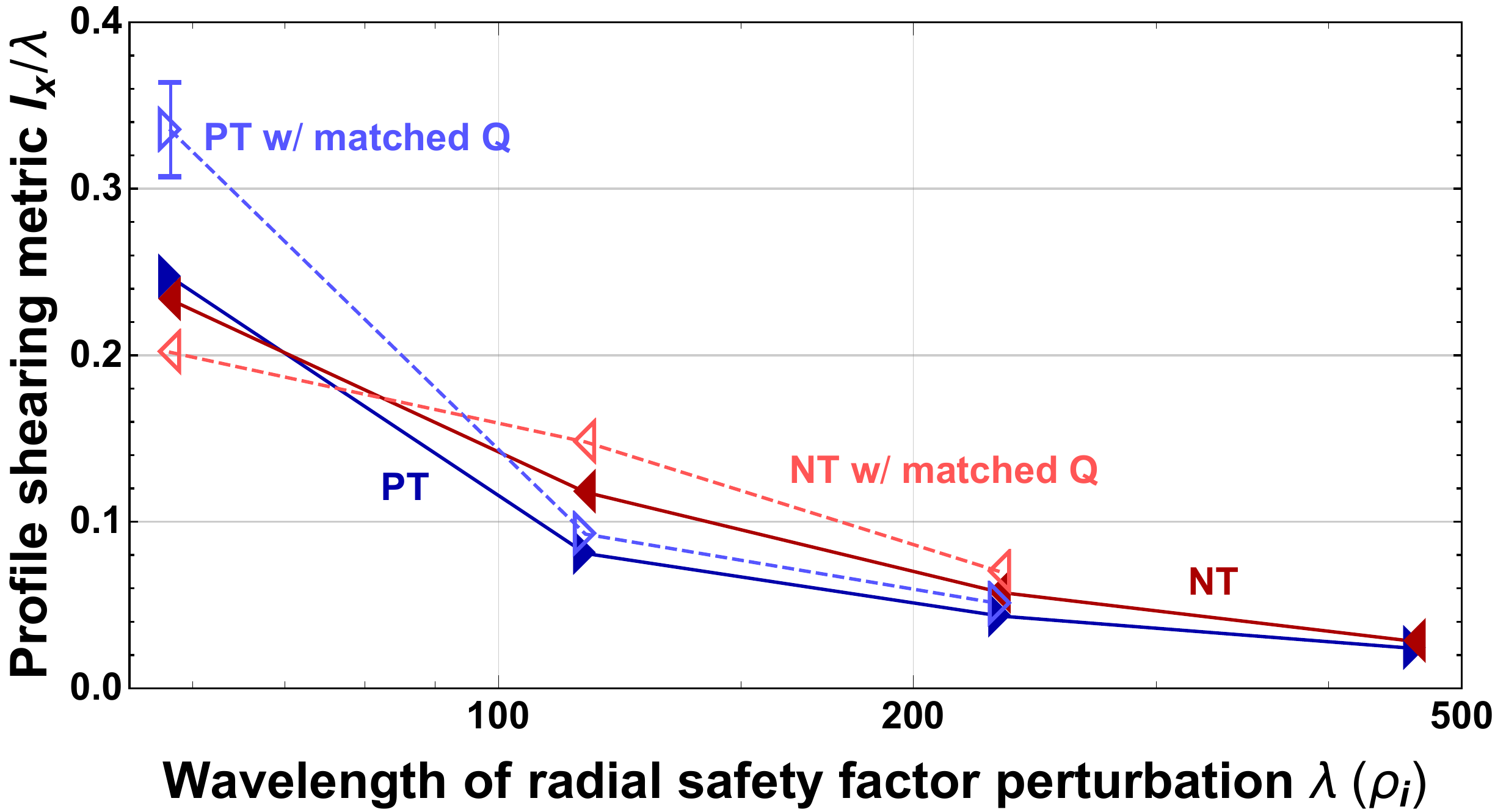}
	\end{center}
	\caption{The (a) linear growth rate $\gamma$, (b) total nonlinear heat flux $Q$, and (c) profile shearing strength metric $l_{x}/\lambda$ as a function of the wavelength $\lambda$ of a sinusoidal safety factor modification for PT with $R_{0}/L_{Ts} = 9$ (solid dark blue, filled right-pointing triangles), NT with $R_{0}/L_{Ts} = 9$ (solid dark red, filled left-pointing triangles), PT with $R_{0}/L_{Ts} = 6$ (dashed light blue, empty right-pointing triangles), and NT with $R_{0}/L_{Ts} = 12.9$ in (a) or $R_{0}/L_{Ts} = 7.2$ in (b) (dashed light red, empty left-pointing triangles). The horizontal grid lines in (a) and (b) indicate the result for uniform magnetic shear.}
	\label{fig:TildeScans}
\end{figure}

Figure \ref{fig:TildeScans}(a) shows the results of a linear scan. We see that at short wavelength the growth rate is strongly reduced, while it converges to the uniform $\hat{s}$ result at long wavelength. This makes sense. Since the amplitude of the radial variation in magnetic shear is proportional to $d \tilde{q}/dr \sim \tilde{q}_{n}^{S} / \lambda$, at short wavelength the variation in the magnetic shear across the domain is very strong, while at sufficiently long wavelength the variation in the shear becomes negligible. We also see that profile shearing is stabilizing, which is consistent with past work using global simulations \cite{McMillanSystemSizeScaling2010, MerloNegTriGlobal2021, WaltzProfileShearing2002}. Figure \ref{fig:TildeScans}(a) shows data for PT and NT equilibria that have the same background gradients of density and temperature. However, due to the stabilizing effect of its geometry, the linear growth rate is much lower for NT. To control for the change in the linear drive, we ran two additional cases in which we modified the strength of the background temperature gradients in order to match the linear growth rates for PT and NT at large $\lambda$. Surprisingly, we see that the NT cases converge more quickly as $\lambda \rightarrow \infty$, regardless of the strength of the linear drive. This indicates that, in this linear study, global effects are {\it less} important for negative triangularity, which is the opposite result of the nonlinear study in \cite{MerloNegTriGlobal2021}. However, linear studies have important limitations. Individual linear results can be idiosyncratic as the fastest-growing mode can be localized to particular radial regions. Additionally, it does not include any of the physics of turbulent saturation, which can be substantially different between PT and NT \cite{DuffNTsaturation2022}.

Thus, we performed an analogous nonlinear study, shown in figure \ref{fig:TildeScans}(b). Because the total heat fluxes in the standard PT and NT cases were fairly similar, we investigated the impact of the drive by reducing the gradients in both cases, but such that their total heat fluxes roughly matched one another's. We see that {\it all} of the cases converge quite similarly with $\lambda$ --- neither the plasma shape, nor the strength of the drive appear to significantly influence the impact of the profile shearing effect. To understand this, we further analyzed the data by looking at the metric $l_{x} / \lambda$, where $l_{x}$ is a measure of the radial size of the turbulent eddies and $\lambda$ is radial wavelength of the profile shearing. One would expect the impact of profile shearing to decrease as $l_{x} / \lambda$ gets smaller because eddies can't be stabilized by profile shearing if they aren't large enough to perceive the profile shearing. Figure \ref{fig:TildeScans}(c) shows this metric computed for each simulation (taking $l_{x}$ to be the e-folding eddy diameter of the radial two-point correlation function of the non-zonal $\phi$). We find that $l_{x} / \lambda$ is relatively insensitive to the plasma shape and, more surprisingly, to the drive as well. Since stronger turbulent drive causes higher heat flux, we expected that this would have to result from larger eddies. However, while the simulations with stronger drive did have larger heat flux, the average radial size of the eddies remained approximately unchanged. In hindsight, this might still be consistent with scaling arguments that predict $l_{x} \sim R_{0}/L_{Ts}$ \cite{BarnesCriticalBalance2011}, since $R_{0}/L_{Ts}$ is only varying by a relatively small amount (i.e. 25\% or 50\%). Granted this surprise, however, the results between figures \ref{fig:TildeScans}(b) and (c) are consistent. The metric $l_{x} / \lambda$ is relatively insensitive to plasma shape and drive while varying much more substantially with $\lambda$, which is also true of the total heat flux.

Nevertheless, it is important to note that the irrelevance of the plasma shape to profile shearing differs from the global nonlinear results of \cite{MerloNegTriGlobal2021}. The reasons behind this are not obvious and a deeper understanding would require additional simulations of the TCV equilibria (both global and local with non-uniform magnetic shear). This is outside the scope of this work. Still, there are several differences between the studies that could be important. First, this study holds the magnitude of the profile shearing constant, while \cite{MerloNegTriGlobal2021} uses measured TCV plasma profiles that are substantially different between PT and NT. Perhaps the NT TCV profiles happen to have stronger profile shearing. Additionally, this study and  \cite{MerloNegTriGlobal2021} use equilibria with different physical parameters and hold different quantities constant in the PT-NT comparison. Perhaps in some regions of parameter space switching from PT to NT changes the eddy size substantially, while in others it doesn't. Other more technical differences between the two studies include the physical effects included in the simulation (e.g. \cite{MerloNegTriGlobal2021} includes collisions, impurities, and electromagnetic effects), the model used (non-uniform magnetic shear versus global simulations), and the geometry specification (idealized analytic Miller versus numerical TCV). Regardless of the reason, this study indicates that global effects are {\it not} universally more important in tokamaks with NT plasma shaping.

%%===================================================%
%%===================================================%
\section{Conclusions}
\label{sec:conclusions}
%%===================================================%
%%===================================================%

In this work, we have rigorously derived a reasonable and internally consistent gyrokinetic model that includes ion gyroradius-scale variation in the magnetic shear profile. This was done by adding an ECCD-inspired current drive source to the Fokker-Planck equation, performing the standard gyrokinetic expansion, and making a subsidiary asymptotic expansion to separate the velocity scale of the fast electrons driven by ECCD from the typical thermal velocity of the electrons. Using this, we modeled a current drive source that varies on a radial scale of tens to hundreds of gyroradii, yet still asymptotically smaller than the tokamak minor radius. We found that the effect of such a source can be made identical to locally modifying the radial profile of the safety factor. The derivation produces a gyrokinetic model with new terms that adjust the ion and electron parallel streaming to be along the modified field lines. This functionality was implemented in the gyrokinetic code GENE (retaining the typical Fourier-space representation in the perpendicular spatial directions) and was successfully benchmarked against analytic results as well as standard nonlinear flux tube simulations. This functionality enables one to add arbitrary periodic radial variation to the magnetic shear profile within the flux tube. However, it does cause the coordinate system to depart from being exactly field-aligned. As a result, if the amplitude of the variation becomes too large, properly resolving the turbulence requires one to increase the parallel resolution proportionally.

As an additional benefit of the model, the non-uniform magnetic shear causes the turbulent transport characteristics to vary within the domain. Thus, since there are no sources of energy, particles, or momentum, the turbulence creates steady zonal perturbations that adjust the corresponding background gradients such that all fluxes are constant across the domain. This means that these simulations naturally include the profile shearing in temperature, density, and flow that are self-consistent with the imposed profile shearing in the safety factor. To illustrate possible applications, we used the modified GENE code to study the importance of profile shearing in equilibria with positive and negative triangularity. Using nonlinear simulations, we found little difference between the two.

In the future, a flux tube with non-uniform magnetic shear, as developed in this paper, could have a number of applications. First, it could enable efficient and reliable simulations of reversed shear safety factor profiles, which is particularly relevant for studying internal transport barriers. Second, non-uniform magnetic shear may reduce the computational cost of simulating very low but finite values of magnetic shear. One can create wide radial regions of very low shear within the flux tube, while still having a moderate value of magnetic shear on average across the box. Thus, unlike standard low shear simulations, the radial size of the flux tube will not be constrained to be very large by the box discretization condition \cite{BeerBallooningCoordinates1995}. Third, using electromagnetic simulations, one can study how the turbulence self-generates magnetic fields in reaction to the externally imposed safety factor variation. In other words, at finite $\beta$, to what degree can the plasma cancel out the non-uniform safety factor profile? This may be useful in understanding how, for example, turbulence broadens the current driven by ECCD. One could create a safety factor profile with a single sharp spike and see how it is broadened by the magnetic fields generated by the turbulence. Fourth, one could search for non-uniform magnetic shear profiles that create large variation in the self-consistent temperature and density profiles, and then use them to directly study the impact of temperature and density profile shearing.

\ack

The authors would like to thank Ian Abel, Jean Cazabonne, Peter Donnel, Alessandro Geraldini, and Arnas Vol\v{c}okas for useful discussions pertaining to this work. 
This work has been carried out within the framework of the EUROfusion Consortium, funded by the European Union via the Euratom Research and Training Programme (Grant Agreement No 101052200 --- EUROfusion). Views and opinions expressed are however those of the author(s) only and do not necessarily reflect those of the European Union or the European Commission. Neither the European Union nor the European Commission can be held responsible for them.
We acknowledge the CINECA award under the ISCRA initiative, for the availability of high performance computing resources and support. 
This work was supported by a grant from the Swiss National Supercomputing Centre (CSCS) under project ID 1050 and 1097. 
This work was carried out using the JFRS-1 supercomputer system at Computational Simulation Centre of International Fusion Energy Research Centre (IFERC-CSC) in Rokkasho Fusion Institute of QST (Aomori, Japan).

\appendix

%===================================================%
%===================================================%
\section{Derivation of the modified electron kinetic equation}
\label{app:elecKinEq}
%===================================================%
%===================================================%

In this appendix we will rigorously derive the kinetic equation governing electron dynamics in the presence of the ion gyroradius-scale current drive source, using the multi-scale asymptotic analysis described in section \ref{sec:deriv}. Since we plan to model ion scale turbulence, we will start by taking the drift kinetic limit (i.e. $\vec{\rho}_{e} \ll \vec{x} \sim \vec{X}$) for simplicity. Even the high velocity electron tail can be treated as drift kinetic because we have assumed that $v_{f} \ll (m_{i}/m_{e}) v_{thi}$. In this limit, the electron kinetic equation is
\begin{align}
\frac{\partial h_{e}}{\partial t} &+ v_{||} \hat{b} \cdot \Nabla h_{e} + \left( \vec{c}_{\kappa} v_{||}^{2} + \vec{c}_{\nabla B} \mu \right) \cdot \Nabla h_{e} + c_{a||} \mu \frac{\partial h_{e}}{\partial v_{||}} - \sum_{s} C^{L}_{e s} \label{eq:DKeqElec} \\
&- \frac{1}{B} \left( \Nabla h_{e} \times \Nabla \chi \right) \cdot \hat{b} = \frac{Z_{e} e F_{Me}}{T_{e}} \frac{\partial \chi}{\partial t} + \frac{1}{B} \left( \Nabla F_{Me} \times \Nabla \chi \right) \cdot \hat{b} + \tilde{S}^{Ip}_{e} , \nonumber
\end{align}
where the generalized potential has become $\langle \chi \rangle_{\varphi} = \chi = \phi - v_{||} A_{||} + m_{e} \mu / (Z_{e} e) \delta B_{||}$ for electrons. Substituting $v_{||} \rightarrow v_{||} + v_{||f}$ and $\mu \rightarrow \mu + \mu_{f}$ , then expanding to lowest order in $\epsilon \ll 1$ gives the $O( \epsilon^{-2} h_{e} v_{the}/a )$ equation
\begin{align}
\vec{c}_{\kappa} v_{||f}^{2} \cdot \Nabla h_{e0} = 0 . \label{eq:DKeqElecNeg2}
\end{align}
Note that no turbulent drive terms ever appear at the fast velocity scales because the Maxwellian distribution function becomes exponentially small at high velocities. Given that the perturbed distribution function has no spatially uniform contribution, equation \refEq{eq:DKeqElecNeg2} implies that the electron distribution function can only be non-zero when $v_{||f} = 0$ (except at poloidal locations where components of $\vec{c}_{\kappa}$ vanish, which are treated in \ref{app:radialDrift}). Since equation \refEq{eq:DKeqElecNeg2} gives no information about the distribution function when $v_{||f} = 0$, we define $h_{ej}^{||} (v_{||}, \mu, \mu_{f}) \equiv h_{ej}  (v_{||}, \mu, v_{||f} = 0, \mu_{f})$ for any asymptotic order $j$ in the $\epsilon \ll 1$ expansion. This allows us to write
\begin{align}
h_{e0}  (v_{||}, \mu, v_{||f}, \mu_{f}) = \left\{
\begin{array}{ll}
h_{e0}^{||}  (v_{||}, \mu, \mu_{f}) & v_{||f} = 0 \label{eq:he0vpar} \\
0 & \text{else}
\end{array}
\right. .
\end{align}

Expanding to $O( \epsilon^{-1} h_{e} v_{the}/a)$ and employing equation \refEq{eq:he0vpar} gives
\begin{align}
\vec{c}_{\kappa} v_{||f}^{2} \cdot \Nabla h_{e1} &+ \vec{c}_{\nabla B} \mu_{f} \cdot \Nabla h_{e0}^{||} + c_{a||} \mu_{f} \frac{\partial h_{e0}^{||}}{\partial v_{||}} - \frac{1}{B} \left( \Nabla h_{e0}^{||} \times \Nabla \left( \frac{m_{e} \mu_{f}}{Z_{e} e} \delta B_{||0} \right) \right) \cdot \hat{b} = 0 . \label{eq:DKeqElecNeg1}
\end{align}
Evaluating this at $v_{||f} = 0$ and noting that there are no drive terms, we see that $h_{e0}^{||} =0$ unless $\mu_{f} = 0$. Combining this result with equation \refEq{eq:he0vpar} demonstrates that
\begin{align}
h_{e0}  (v_{||}, \mu, v_{||f}, \mu_{f}) = \left\{
\begin{array}{ll}
h_{e0}^{th}  (v_{||}, \mu) & v_{||f} = \mu_{f} = 0 \label{eq:he0vth} \\
0 & \text{else}
\end{array}
\right. ,
\end{align}
where we define $h_{ej}^{th} (v_{||}, \mu) \equiv h_{ej}^{||}  (v_{||}, \mu, \mu_{f} = 0)$ at any asymptotic order $j$. Here $h_{ej}^{th}$ is the electron distribution function at thermal velocities, which is what standard kinetic codes calculate. In other words, the lowest order electron distribution function has no high velocity tail and can only have activity at thermal speeds. This is the same result as we obtained earlier for the ion distribution function and is intuitive given that the source $\tilde{S}^{Ip}_{e}$ has yet to appear. Substituting equation \refEq{eq:he0vth} into equation \refEq{eq:DKeqElecNeg1} gives
\begin{align}
\vec{c}_{\kappa} v_{||f}^{2} \cdot \Nabla h_{e1} = 0 . \label{eq:DKeqElecNeg1vf}
\end{align}
Therefore, as at lowest order we find
\begin{align}
h_{e1}  (v_{||}, \mu, v_{||f}, \mu_{f}) = \left\{
\begin{array}{ll}
h_{e1}^{||}  (v_{||}, \mu, \mu_{f}) & v_{||f} = 0 \label{eq:he1vpar} \\
0 & \text{else}
\end{array}
\right. .
\end{align}

Going to $O(h_{e} v_{the}/a)$ in our expansion of equation \refEq{eq:DKeqElec} and then evaluating the result at $v_{||f} = \mu_{f} = 0$ gives
\begin{align}
\frac{\partial h_{e0}^{th}}{\partial t} &+ v_{||} \hat{b} \cdot \Nabla h_{e0}^{th} + \left( \vec{c}_{\kappa} v_{||}^{2} + \vec{c}_{\nabla B} \mu \right) \cdot \Nabla h_{e0}^{th} + c_{a||} \mu \frac{\partial h_{e0}^{th}}{\partial v_{||}} - \sum_{s} C^{L}_{e s 0} \label{eq:DKeqElec0vth} \\
&- \frac{1}{B} \left( \Nabla h_{e0}^{th} \times \Nabla \chi_{0} \right) \cdot \hat{b} = \frac{Z_{e} e F_{Me}}{T_{e}} \frac{\partial \chi_{0}}{\partial t} + \frac{1}{B} \left( \Nabla F_{Me} \times \Nabla \chi_{0} \right) \cdot \hat{b} , \nonumber
\end{align}
where $\chi_{0} = \phi_{0} - v_{||} A_{||0} + m_{e} \mu / (Z_{e} e) \delta B_{||0}$. This equation determines the lowest order electron distribution function at thermal velocities $h_{e0}^{th}$ and has an identical {\it form} to the standard electron drift kinetic equation. Substituting equations \refEq{eq:he0vth}, \refEq{eq:he1vpar}, and \refEq{eq:DKeqElec0vth} into the $O(h_{e} v_{the}/a)$ equation at $v_{||f}=0$ gives
\begin{align}
\vec{c}_{\nabla B} \mu_{f} \cdot \Nabla h_{e1}^{||} + c_{a||} \mu_{f} \frac{\partial h_{e1}^{||}}{\partial v_{||}} - \frac{1}{B} \left( \Nabla h_{e1}^{||} \times \Nabla \left( \frac{m_{e} \mu_{f}}{Z_{e} e} \delta B_{||0} \right) \right) \cdot \hat{b} = 0 . \label{eq:DKeqElec0vpar}
\end{align}
Thus, since there are still no drive terms, we see that $h_{e1}^{||} =0$ unless $\mu_{f} = 0$. Combining this result with equation \refEq{eq:he1vpar} demonstrates that
\begin{align}
h_{e1}  (v_{||}, \mu, v_{||f}, \mu_{f}) = \left\{
\begin{array}{ll}
h_{e1}^{th}  (v_{||}, \mu) & v_{||f} = \mu_{f} = 0 \label{eq:he1vth} \\
0 & \text{else}
\end{array}
\right. .
\end{align}
Substituting equations \refEq{eq:he0vth}, \refEq{eq:DKeqElec0vth}, and \refEq{eq:he1vth} into the $O(h_{e} v_{the}/a)$ equation gives
\begin{align}
\vec{c}_{\kappa} v_{||f}^{2} \cdot \Nabla h_{e2} = 0 , \label{eq:DKeqElec0}
\end{align}
so we know that
\begin{align}
h_{e2}  (v_{||}, \mu, v_{||f}, \mu_{f}) = \left\{
\begin{array}{ll}
h_{e2}^{||}  (v_{||}, \mu, \mu_{f}) & v_{||f} = 0 \label{eq:he2vpar} \\
0 & \text{else}
\end{array}
\right. .
\end{align}

Lastly, the $O(\epsilon h_{e} v_{the}/a)$ electron drift kinetic equation is fairly lengthy, but finally features the lowest order current drive source $\tilde{S}^{Ip}_{e0}$. Evaluating this equation at $v_{||f} = \mu_{f} = 0$ gives an equation that determines $h_{e1}^{th}$, but this quantity will not be needed. Substituting this equation into the $O(\epsilon h_{e} v_{the}/a)$ electron drift kinetic equation evaluated at $v_{||f} = 0$ produces
\begin{align}
\vec{c}_{\nabla B} \mu_{f} \cdot \Nabla h_{e2}^{||} + c_{a||} \mu_{f} \frac{\partial h_{e2}^{||}}{\partial v_{||}} - \frac{1}{B} \left( \Nabla h_{e2}^{||} \times \Nabla \left( \frac{m_{e} \mu_{f}}{Z_{e} e} \delta B_{||0} \right) \right) \cdot \hat{b} = 0 . \label{eq:DKeqElec1vpar}
\end{align}
Importantly, we used the assumption that $\tilde{S}^{Ip}_{e} ( v_{||f} = 0, \mu_{f} ) = 0$ to prevent the current drive source from appearing on the right-hand side. Equation \refEq{eq:DKeqElec1vpar} demonstrates that 
\begin{align}
h_{e2}  (v_{||}, \mu, v_{||f}, \mu_{f}) = \left\{
\begin{array}{ll}
h_{e2}^{th}  (v_{||}, \mu) & v_{||f} = \mu_{f} = 0 \label{eq:he2vth} \\
0 & \text{else}
\end{array}
\right.
\end{align}
has no high velocity electron tail. Finally, substituting this all into the $O(\epsilon h_{e} v_{the}/a)$ electron drift kinetic equation produces
\begin{align}
\vec{c}_{\kappa} v_{||f}^{2} \cdot \Nabla h_{e3} = \tilde{S}^{Ip}_{e0} ( v_{||f}, \mu_{f} ) , \label{eq:DKeqElec1}
\end{align}
which governs the lowest order non-zero high velocity electron tail. We will label the third-order distribution function $h_{e3}^{f} \equiv h_{e3}$ with a superscript $f$ as a reminder that it contains activity at fast velocity scales. Given that we can assume that the source doesn't vary in the binormal direction within the flux tube and the parallel derivative is small due to the anisotropy of the turbulence, we can finally calculate the fast electron tail to be given by equation \refEq{eq:he3vf}.

%===================================================%
%===================================================%
\section{Derivation of the modified field equations}
\label{app:fieldEqs}
%===================================================%
%===================================================%

In this appendix we will rigorously carry out the multi-scale asymptotic expansion of the quasineutrality equation and Ampere's law, as described in section \ref{sec:deriv}. Given that we have taken the drift kinetic limit for electrons, equations \refEq{eq:fieldPhi} through \refEq{eq:fieldBpar} become
\begin{align}
\phi &= \left( \sum_{s} \frac{Z_{s}^{2} e^{2} n_{s}}{T_{s}} \right)^{-1} \left( \sum_{i} Z_{i} e B \int_{-\infty}^{\infty} dv_{||} \int_{0}^{\infty} d \mu \left. \oint_{0}^{2 \pi} \right|_{\vec{x}} d \varphi ~ h_{i} \left( \vec{x} - \vec{\rho}_{i} \right) \right. \label{eq:fieldPhiDrift} \\
& \left. + 2 \pi Z_{e} e B \int_{-\infty}^{\infty} dv_{||} \int_{0}^{\infty} d \mu ~ h_{e} \right) \nonumber \\
- \nabla_{\perp}^{2} A_{||} &= \mu_{0} \left( \sum_{i} Z_{i} e B \int_{-\infty}^{\infty} dv_{||} \int_{0}^{\infty} d \mu \left. \oint_{0}^{2 \pi} \right|_{\vec{x}} d \varphi ~ v_{||} h_{i} \left( \vec{x} - \vec{\rho}_{i} \right) \right. \label{eq:fieldAparDrift} \\
& \left. + 2 \pi Z_{e} e B \int_{-\infty}^{\infty} dv_{||} \int_{0}^{\infty} d \mu ~ v_{||} h_{e} \right) \nonumber \\
\Nabla \delta B_{||} \times \hat{b} &= \mu_{0} \left( \sum_{i} Z_{i} e B \int_{-\infty}^{\infty} dv_{||} \int_{0}^{\infty} d \mu \left. \oint_{0}^{2 \pi} \right|_{\vec{x}} d \varphi ~ \vec{v}_{\perp} h_{i} \left( \vec{x} - \vec{\rho}_{i} \right) \right. \label{eq:fieldBparDrift} \\
& \left. - 2 \pi m_{e} B \int_{-\infty}^{\infty} dv_{||} \int_{0}^{\infty} d \mu ~ \mu \Nabla h_{e} \times \hat{b} \right) . \nonumber
\end{align}
Expanding each of these to lowest order in $\epsilon \ll 1$ and using equations \refEq{eq:he3vf}, \refEq{eq:he0vth}, \refEq{eq:he1vth}, and \refEq{eq:he2vth} gives
\begin{align}
\phi_{0} &= \left( \sum_{s} \frac{Z_{s}^{2} e^{2} n_{s}}{T_{s}} \right)^{-1} \left( \sum_{i} Z_{i} e B \int_{-\infty}^{\infty} dv_{||} \int_{0}^{\infty} d \mu \left. \oint_{0}^{2 \pi} \right|_{\vec{x}} d \varphi ~ h_{i0} \left( \vec{x} - \vec{\rho}_{i} \right) \right. \label{eq:fieldPhiDriftExp} \\
& \left. + 2 \pi Z_{e} e B \int_{-\infty}^{\infty} dv_{||} \int_{0}^{\infty} d \mu ~ h_{e0}^{th} \right) \nonumber \\
- \nabla_{\perp}^{2} A_{|| 0} &= \mu_{0} \left( \sum_{i} Z_{i} e B \int_{-\infty}^{\infty} dv_{||} \int_{0}^{\infty} d \mu \left. \oint_{0}^{2 \pi} \right|_{\vec{x}} d \varphi ~ v_{||} h_{i0} \left( \vec{x} - \vec{\rho}_{i} \right) \right. \label{eq:fieldAparDriftExp} \\
& \left. + 2 \pi Z_{e} e B \int_{-\infty}^{\infty} dv_{||} \int_{0}^{\infty} d \mu v_{||} h_{e0}^{th} + 2 \pi Z_{e} e B \int_{-\infty}^{\infty} dv_{||f} \int_{0}^{\infty} d \mu_{f} ~ v_{||f} h_{e3}^{f} \right) \nonumber \\
\Nabla \delta B_{|| 0} \times \hat{b} &= \mu_{0} \left( \sum_{i} Z_{i} e B \int_{-\infty}^{\infty} dv_{||} \int_{0}^{\infty} d \mu \left. \oint_{0}^{2 \pi} \right|_{\vec{x}} d \varphi ~ \vec{v}_{\perp} h_{i0} \left( \vec{x} - \vec{\rho}_{i} \right) \right. \label{eq:fieldBparDriftExp} \\
& \left. - 2 \pi m_{e} B \int_{-\infty}^{\infty} dv_{||} \int_{0}^{\infty} d \mu \mu \Nabla h_{e0}^{th} \times \hat{b} - 2 \pi m_{e} B \int_{-\infty}^{\infty} dv_{||f} \int_{0}^{\infty} d \mu_{f} ~ \mu_{f} \Nabla h_{e3}^{f} \times \hat{b} \right) . \nonumber
\end{align}
Here, in both the parallel and perpendicular components of Ampere's law, we stress the appearance of a new term arising from the fast electron tail created by the current drive source. Even though the size of the fast electron distribution function is very small, its high characteristic velocity causes it to carry electric current that competes with the thermal contribution. Given that equations \refEq{eq:fieldAparDriftExp} and \refEq{eq:fieldBparDriftExp} are linear in $A_{|| 0}$ and $\delta B_{|| 0}$, we can choose to divide the perturbed magnetic field into the portions arising from the thermal distribution and from the fast electron tail according to $A_{|| 0} = A_{|| 0}^{th} + A_{|| 0}^{f}$ and $\delta B_{|| 0} = \delta B_{|| 0}^{th} + \delta B_{|| 0}^{f}$. These fields are defined to satisfy
\begin{align}
- \nabla_{\perp}^{2} A_{|| 0}^{th} &= \mu_{0} \left( \sum_{i} Z_{i} e B \int_{-\infty}^{\infty} dv_{||} \int_{0}^{\infty} d \mu \left. \oint_{0}^{2 \pi} \right|_{\vec{x}} d \varphi ~ v_{||} h_{i0} \left( \vec{x} - \vec{\rho}_{i} \right) \right. \label{eq:fieldAparDriftExpThermal} \\
& \left. + 2 \pi Z_{e} e B \int_{-\infty}^{\infty} dv_{||} \int_{0}^{\infty} d \mu ~ v_{||} h_{e0}^{th} \right) \nonumber \\
\Nabla \delta B_{|| 0}^{th} \times \hat{b} &= \mu_{0} \left( \sum_{i} Z_{i} e B \int_{-\infty}^{\infty} dv_{||} \int_{0}^{\infty} d \mu \left. \oint_{0}^{2 \pi} \right|_{\vec{x}} d \varphi ~ \vec{v}_{\perp} h_{i0} \left( \vec{x} - \vec{\rho}_{i} \right) \right. \label{eq:fieldBparDriftExpThermal} \\
& \left. - 2 \pi m_{e} B \int_{-\infty}^{\infty} dv_{||} \int_{0}^{\infty} d \mu ~ \mu \Nabla h_{e0}^{th} \times \hat{b} \right) \nonumber
\end{align}
as well as equations \refEq{eq:fieldAparDriftExpFast} and \refEq{eq:fieldBparDriftExpFast}. The first two equations are the standard field equations solved by gyrokinetic codes, while equations \refEq{eq:fieldAparDriftExpFast} and \refEq{eq:fieldBparDriftExpFast} are the new contributions to the perturbed perpendicular and parallel magnetic field from the current drive source. Analogously, we can write the generalized potential appearing in the electron {\it drift} kinetic equation as
\begin{align}
\chi_{0} = \chi_{0}^{th} + \chi_{0}^{f} , \label{eq:chiForm}
\end{align}
where $\chi_{0}^{th} = \phi_{0} - v_{||} A_{|| 0}^{th} + m_{e} \mu/ ( Z_{e} e ) \delta B_{|| 0}^{th}$ and $\chi_{0}^{f} = - v_{||} A_{|| 0}^{f} + m_{e} \mu/ ( Z_{e} e ) \delta B_{|| 0}^{f}$, and the generalized potential appearing in the ion {\it gyro}kinetic equation as
\begin{align}
\langle \chi_{0} \rangle_{\varphi} = \langle \chi_{0} \rangle_{\varphi}^{th} + \langle \chi_{0} \rangle_{\varphi}^{f} , \label{eq:chiFormIon}
\end{align}
where $\langle \chi_{0} \rangle_{\varphi}^{th} \equiv \langle \phi_{0} \rangle_{\varphi} - v_{||} \langle A_{|| 0}^{th} \rangle_{\varphi} - \langle \vec{v}_{\perp} \cdot \vec{A}_{\perp 0}^{th} \rangle_{\varphi}$ and $\langle \chi_{0} \rangle_{\varphi}^{f} \equiv - v_{||} \langle A_{|| 0}^{f} \rangle_{\varphi} - \langle \vec{v}_{\perp} \cdot \vec{A}_{\perp 0}^{f} \rangle_{\varphi}$.

From equations \refEq{eq:he3vf}, \refEq{eq:fieldAparDriftExpFast}, \refEq{eq:fieldBparDriftExpFast}, and the form of $\chi_{0}^{f}$ we see that, since $\tilde{S}^{Ip}_{e}$ is independent of $t$ and $\alpha$, so is $A_{|| 0}^{f}$, $\delta B_{|| 0}^{f}$, and $\chi_{0}^{f}$. Thus, in the kinetic equations for ions and electrons (i.e. equations \refEq{eq:GKeqIons} and \refEq{eq:DKeqElec0vth}), $\chi_{0}^{f}$ only survives through the nonlinear term and is eliminated by the time derivative and the turbulent drive term. Substituting equation \refEq{eq:chiForm} and the forms of $\chi_{0}^{f}$ and $\vec{c}_{\nabla B}$ into equation \refEq{eq:DKeqElec0vth}, we see that the electron dynamics are governed by equation \refEq{eq:finalElecEq}. Similarly, by substituting equation \refEq{eq:chiFormIon} into equation \refEq{eq:GKeqIons}, we see that the ion dynamics follow equation \refEq{eq:finalIonEq}.

%===================================================%
%===================================================%
\section{Special case of vanishing magnetic drift components}
\label{app:radialDrift}
%===================================================%
%===================================================%

Poloidal locations where components of the magnetic drifts vanish can create singularities in the derivation presented in section \ref{sec:deriv} (e.g. equation \refEq{eq:he3vf} diverges where $\vec{c}_{\kappa} \cdot \Nabla r = 0$). The purpose of this appendix is to show that, {\it even at these locations}, a current drive source $\tilde{S}^{Ip}_{e}$ can always be found to create the pure form of the safety factor modification assumed by equation \refEq{eq:alphaTildeDef} (i.e. a modification to the magnetic field that preserves the straight-field line coordinate $\theta$ and corresponds to arbitrary long-wavelength radial variation of $\tilde{q} (r)$). Strictly-speaking, doing this is necessary to justify specifying $\tilde{q} (r)$ as an input to a gyrokinetic code, instead of having to use $\tilde{S}^{Ip}_{e}$.

In practice, a vanishing {\it binormal} component does not cause any issues because the current drive source $\tilde{S}^{Ip}_{e}$ is uniform in this direction, as is the fast electron tail that it drives. However, poloidal locations where $\vec{c}_{\kappa} \cdot \Nabla r = 0$ clearly require special treatment. This occurs only where $\partial B / \partial \theta = 0$ (typically the inboard and outboard midplanes), which also implies that $\vec{c}_{\nabla B} \cdot \Nabla r = c_{a||} = 0$.

When the radial magnetic drifts vanish, the dominant term balancing the current drive source becomes parallel streaming, which is one order weaker in $\epsilon \ll 1$ than the curvature drift term. Thus, we will order $\tilde{S}^{Ip}_{e} \sim \tilde{S}^{Ip}_{e1} \sim \epsilon^{2} h_{e} v_{the}/a$ to be one order smaller at these locations, which we will see results in the size of the high velocity electron tail remaining the same. Where $\vec{c}_{\kappa} \cdot \Nabla r = \vec{c}_{\nabla B} \cdot \Nabla r = c_{a||} = 0$, the $O( \epsilon^{-2} h_{e} v_{the}/a )$ electron drift kinetic equation (i.e. equation \refEq{eq:DKeqElecNeg2}) becomes
\begin{align}
  \vec{c}_{\kappa} v_{||f}^{2} \cdot \Nabla \alpha \frac{\partial h_{e0}}{\partial \alpha} = 0 , \label{eq:DKeqElecNeg2NoDrift}
\end{align}
showing that $h_{e0}$ must be constant in $\alpha$ when $v_{||f} \neq 0$. Using this result in the $O( \epsilon^{-1} h_{e} v_{the}/a )$ equation gives
\begin{align}
  v_{||f} \hat{b} \cdot \Nabla h_{e0} &+ \vec{c}_{\kappa} v_{||f}^{2} \cdot \Nabla \alpha \frac{\partial h_{e1}}{\partial \alpha} + \vec{c}_{\nabla B} \mu_{f} \cdot \Nabla \alpha \frac{\partial h_{e0}}{\partial \alpha} \label{eq:DKeqElecNeg1NoDrift} \\
  &+ \frac{v_{||f}}{B} \frac{\partial h_{e0}}{\partial r} \frac{\partial A_{||0}}{\partial \alpha} \left( \Nabla r \times \Nabla \alpha  \right) \cdot \hat{b} - \frac{m_{e} \mu_{f}}{Z_{e} e B} \left( \Nabla h_{e0} \times \Nabla \delta B_{||0} \right) \cdot \hat{b} = 0 . \nonumber
\end{align}
Evaluating this at $v_{||f} = 0$ and noting the lack of drive terms, we see that $h_{e0} = 0$ when $\mu_{f} \neq 0$. Substituting this result into equation \refEq{eq:DKeqElecNeg1NoDrift}, averaging over $\alpha$ with $(1/L_{y}) \oint_{0}^{L_{y}} d \alpha ( \ldots )$, and using the binormal periodicity of the flux tube, we find that
\begin{align}
  v_{||f} \hat{b} \cdot \Nabla \theta \frac{\partial h_{e0}}{\partial \theta} = 0 . \label{eq:DKeqElecNeg1NoDriftAvg}
\end{align}
We can then enforce continuity in $\theta$, using the solution of equation \refEq{eq:he0vth} at high velocities, to demonstrate that $h_{e0} = 0$ even at locations where $\vec{c}_{\kappa} \cdot \Nabla r = 0$. Substituting this result into equation \refEq{eq:DKeqElecNeg1NoDrift} shows that $h_{e1}$ must be constant in $\alpha$, analogously to equation \refEq{eq:DKeqElecNeg2NoDrift}. This process can be repeated order by order, noting that the equations at thermal velocity scales (e.g. equation \refEq{eq:DKeqElec0vth} and higher order versions) are valid at all poloidal locations and can be employed to remove the thermal terms from an equation. Using this method, one can show that $h_{e1} = h_{e2} = 0$ when $v_{||f} \neq 0$ or $\mu_{f} \neq 0$, and $h_{e3}$ must be constant in $\alpha$ when $v_{||f} \neq 0$. At $O( \epsilon^{2} h_{e} v_{the}/a )$, the current drive source finally enters and we find
\begin{align}
\frac{\partial h_{e3}^{f}}{\partial \theta} = \frac{\tilde{S}^{Ip}_{e1} ( v_{||f}, \mu_{f} )}{v_{||f} \hat{b} \cdot \Nabla \theta} , \label{eq:he3vfNoDrift}
\end{align}
the analog to equation \refEq{eq:he3vf} at poloidal locations where $\vec{c}_{\kappa} \cdot \Nabla r = 0$. Interestingly, this is a constraint on the poloidal derivative of the distribution function, meaning that the distribution function itself is determined by continuity with the solution at neighboring poloidal locations. Thus, we see that equation \refEq{eq:he3vf} still holds where $\vec{c}_{\kappa} \cdot \Nabla r = 0$, where we stress that equation \refEq{eq:he3vf} does not diverge because we have chosen that $\tilde{S}^{Ip}_{e0} \propto \vec{c}_{\kappa} \cdot \Nabla r$ according to equation \refEq{eq:sourcePolVariation}.

To show that a current drive source $\tilde{S}^{Ip}_{e}$ exists for every $\tilde{q} (r)$ modification, we combine equations \refEq{eq:fieldAparDriftExpFast} and \refEq{eq:qTildeAparRelation} to find
\begin{align}
  \frac{\partial^{2} \tilde{q}}{\partial r^{2}} = - \frac{2 \pi \mu_{0} Z_{e} e B}{ (\hat{b} \cdot \Nabla \theta) | \Nabla r |^{2} }  \int_{-\infty}^{\infty} dv_{||f} \int_{0}^{\infty} d \mu_{f} ~v_{||f} \frac{\partial h_{e3}^{f}}{\partial \psi} . \label{eq:qTildehe3Relation}
\end{align}
We must show that $\partial^{2} \tilde{q} / \partial r^{2}$ is continuous with $\theta$ and it is constant with $\theta$ (i.e. its poloidal derivative is zero) across locations with $\vec{c}_{\kappa} \cdot \Nabla r = 0$. Given that we just enforced continuity for $h_{e3}^{f}$ following equation \refEq{eq:he3vfNoDrift}, equation \refEq{eq:qTildehe3Relation} itself is clearly continuous with $\theta$. To demonstrate that equation \refEq{eq:qTildehe3Relation} is constant with $\theta$, we will prove that its poloidal derivative is zero everywhere. Since all other quantities are well-behaved with $\theta$ and we already showed in section \ref{sec:deriv} that $\partial ( \partial^{2} \tilde{q} / \partial r^{2} )/ \partial \theta = 0$ where $\vec{c}_{\kappa} \cdot \Nabla r \neq 0$, we only need to show that $\partial ( \partial h_{e3}^{f} / \partial \psi )/ \partial \theta$ is continuous in $\theta$ across the locations where $\vec{c}_{\kappa} \cdot \Nabla r = 0$. Thus, we can take the radial derivative of equation \refEq{eq:he3vfNoDrift} and equate it to the poloidal derivative of equation \refEq{eq:he3vf} to show that
\begin{align}
  \frac{\partial \tilde{S}^{Ip}_{e1}}{\partial r} = \frac{\hat{b} \cdot \Nabla \theta}{v_{||f}} \frac{\partial}{\partial \theta} \left( \frac{\tilde{S}^{Ip}_{e0}}{\vec{c}_{\kappa} \cdot \Nabla r} \right) , \label{eq:sourceFormNextOrder}
\end{align}
which again does not diverge as we have chosen that $\tilde{S}^{Ip}_{e0} \propto \vec{c}_{\kappa} \cdot \Nabla r$. As long as the next order contribution to the source is chosen according to this equation, the modifications to the magnetic field created by the current drive source will correspond to the pure safety factor modification assumed by the form of equation \refEq{eq:alphaTildeDef}, even at poloidal locations where $\vec{c}_{\kappa} \cdot \Nabla r = 0$. Thus, we are free to specify the safety factor profile $\tilde{q} (r)$ as an input to gyrokinetic codes, rather than the current drive source $\tilde{S}^{Ip}_{e}$.

%===================================================%
%===================================================%
\section*{References}
\bibliographystyle{unsrt}
\bibliography{references.bib}
%===================================================%
%===================================================%

\end{document}